*Article*

# Taking the pulse of COVID-19: A spatiotemporal perspective

**(Latest and Live Animations Version)**

**Chaowei Yang [1,2,*], Dexuan Sha [1,2], Qian Liu [1,2], Yun Li [1,2], Hai Lan [1,3], Weihe Wendy Guan [4], Tao Hu [4], Zhenlong Li [5], Zhiran Zhang [1,6,21], John Hoot Thompson [7], Zifu Wang [1,2], David Wong [2], Shiyang Ruan [2], Manzhu Yu [8], Douglas Richardson [4], Luyao Zhang [9], Ruizhi Hou [10], You Zhou [1,11], Cheng Zhong [1,12], Yifei Tian [1,13], Fayez Beaini [1,14], Kyla Carte [1,2], Colin Flynn [2], Wei Liu [1,15], Dieter Pfoser [2], Shuming Bao [16], Mei Li [17], Haoyuan Zhang [17], Chunbo Liu [18], Jie Jiang [19], Shihong Du [17], Liang Zhao[13], Mingyue Lu [20], Lin Li [21], Huan Zhou [22]**

[1] NSF Spatiotemporal Innovation Center, George Mason Univ., Fairfax, VA 22030, USA; cyang3@gmu.edu (C.Y.); dsha@gmu.edu (D.S.); qliu6@gmu.edu (Q.L.); yli38@gmu.edu (Y.L.);
[2] Department of Geography and GeoInformation Science, George Mason Univ., Fairfax, VA, 22030, USA; dwong2@gmu.edu (D.W.); sruan@gmu.edu (S.R.); kcarte@gmu.edu (K.C.); cflynn8@gmu.edu (C.F.); dpfoser@gmu.edu (D.P.);
[3] Department of Geographical Sciences, University of Maryland, College Park, MD 20742, USA; hlan@terpmail.umd.edu (H.L.);
[4] Center for Geographic Analysis, Harvard University, Cambridge, MA 02138, USA; wguan@g.harvard.edu (W.G.); taohu@g.harvard.edu (T.H.); dbrichardson@fas.harvard.edu (D.R.);
[5] Geoinformation and Big Data Research Lab, Department of Geography, University of South Carolina, Columbia, SC 29208, USA; zhenlong@sc.edu (Z.L.);
[6] Chinese Academy of Surveying and Mapping, Beijing, China;
[7] Cloud, Compute, and Storage Operations, Information Technology Services, George Mason University, Fairfax, VA, 22030, USA, jthomp58@gmu.edu (J.T.)
[8] Department of Geography, Pennsylvania State University, State College, PA 16801, USA; mqy5198@psu.edu (M.Y.)
[9] School of Business Management, East China Normal University, Shanghai, 3663 Zhongshanbei Road, China; sunshineluyao@gmail.com; (L.Z.);
[10] School of Mathematical Sciences, East China Normal University, Shanghai, 3663 Zhongshanbei Road, China; houruizhiecnu@gmail.com; (R.H.);
[11] Department of Psychology, George Mason University, Fairfax, VA, 22030, USA, yzhou21@gmu.edu (Y.Z.)
[12] Department of Computer Science, George Mason University, Fairfax, VA, 22030, USA, czhong2@gmu.edu (C.Z.)
[13] Department of Information Technology, George Mason University, Fairfax, VA, 22030, USA, ytian20@gmu.edu (Y.T.); lzhao@gmu.edu
[14] Department of Biology, George Mason University, Fairfax, VA, 22030, USA, fbeaini@gmu.edu (F.B.)
[15] College of Land Science and Technology, China Agricultural University, Beijing, 100083, China; devilweil@cau.edu.cn (W.L.);
[16] China Data Institute, Ann Arbor, Michigan, USA; sbao@umich.edu (S.B.)
[17] Institute of Remote Sensing and GIS, Peking University, Beijing, China, 100871; limeipku@gmail.com (M.L.); zhanghaoyuan@pku.edu.cn (H.Z.); smilegis@163.com (S.D.)
[18] Aerospace Information Research Institute, Chinese Academy of Sciences, Beijing, 100190, China; liucb@aircas.ac.cn (C.L.)
[19] School of Geomatics and Urban Spatial Informatics, Beijing University of Civil Engineering and Architecture, Beijing, 100044, China; jiangjie_263@263.net (J.J.)
[20] School of Geographical Sciences, Nanjing University of Information Science & Technology, 219 Ningliu Road, Nanjing, Jiangsu 210044, China; lumingyue@nuist.edu.cn (M.L.);
[21] School of Resource and Environmental Sciences, Wuhan University, 129 Luoyu Rd. Wuhan, Hubei 430079, China; zhangzhiran@whu.edu.cn (Z.Z.); lilin@whu.edu.cn (L.L.);
[22] School of Geodesy and Geomatics, Wuhan University, 129 Luoyu Rd. Wuhan, Hubei 430079, China; zhouhuan@whu.edu.cn (H.Z.)

* Correspondence: cyang3@gmu.edu (C.Y.)

**Abstract:** The sudden outbreak of the Coronavirus disease (COVID-19) swept across the world in early 2020, triggering the lockdowns of several billion people across many countries, including China, Spain, India, the U.K., Italy, France, Germany, and most states of the U.S. The transmission of the virus accelerated rapidly with the most confirmed cases in the U.S., and New York City became an epicenter of the pandemic by the end of March. In response to this national and global emergency, the NSF Spatiotemporal Innovation Center brought together a taskforce of international researchers and assembled implemented strategies to rapidly respond to this crisis, for supporting research, saving lives, and protecting the health of global citizens. This perspective paper presents our collective view on the global health emergency and our effort in collecting, analyzing, and sharing relevant data on global policy and government responses, geospatial indicators of the outbreak and evolving forecasts; in developing research capabilities and mitigation measures with global scientists, promoting collaborative research on outbreak dynamics, and reflecting on the dynamic responses from human societies.

---

## 1. Introduction

In December 2019, a viral disease was transmitted in Wuhan, China. Many people showed symptoms of coughing, sneezing, and breathing difficulties, and dozens of people with pneumonia were treated and hospitalized in Wuhan [1]. While early announcements contended that the disease was not contagious and many popular gatherings were held as scheduled in early January, the disease was found to be caused by a novel coronavirus, and was confirmed later that it can spread from person to person. Early analyses estimated the reproduction number R0 of approximately 2.0, meaning that an infected person on average would spread the disease to two others, comparable to SARS and MERS [2]. Consequently, Wuhan city, a metropolitan area with 11 million people, was locked down with strict stay-at-home orders, and the shutdown quickly expanded to the entire province of Hubei [3]. The disease later caused the lockdown of many Chinese provinces in February. Eventually, China had to halt its economy for the entirety of February and most of March to contain the transmission [3]. The World Health Organization (WHO) named the disease COVID-19. Despite China's effective and decisive efforts in containing the outbreak since late January 2020 by locking down the most populous country and the second largest economy in the world, unfortunately, the disease quickly spread around the world[4]. Over twenty-seven thousand people lost their lives in Italy alone by April 28 (https://github.com/stccenter/COVID-19-Data/tree/master/Italy) and its healthcare system almost collapsed. Ventilators had to be moved from older people to save the lives of those younger ones [4].

The U.S. Center for Disease Control and Prevention (CDC) observed the outbreak closely and issued emergency advice in mid-February (https://www.cdc.gov/coronavirus/2019-ncov/). While only a few cases were confirmed in the U.S. in late January, the outbreak was severe in many states by March, with hundreds of thousands of cases confirmed around the country. New York City became the world's epicenter in March and the White House declared a national emergency in response to the fast and widely spreading COVID-19. Many states announced stay-at-home orders (lockdowns) on a household basis. While the daily confirmed cases were on a declining trajectory in 20 states of the U.S. by late April, and many governors planned to reopen the economy, concerns were aired on many different aspects:

- The reproduction number of R0 was recalculated to be approximately 6, three times more contagious than the initial 2.0 [5]. Many leaders around the world were unaware of this contagious magnitude, and thus lost the best (early) time to contain the outbreak.

- Fundamental questions, such as the infectious dose, remain unknown[6]. Furthermore, how long the virus can stay alive in an environment spans a wide range from hours to 7 days, depending on the conditions [6]. Although all evidence shows the virus is natural and transferred from bats or possibly via an intermediate mammal species, the transferability from species to species is unknown, especially from animals to humans [6,7].
- Although most people infected will develop symptoms within 14 days, there are healthy people carrying the COVID-19 virus who show no symptoms (asymptomatic) but are also contagious and could become dangerously contagious sources without self-knowledge [8]. For example, Baggett [9] reported 147 residents out of 408 (36%) residents (at a large Boston shelter) were tested positive for COVID-19 and only a small proportion of them showed coughing (7.5%), shortness of breath (1.4%), and fever (0.7%) among the test-positive individuals. The virus could become deadly when spreading rapidly through these asymptomatic carriers once normal activities resume in metropolitan areas.
- Singapore, thought to be the model for containing the virus from the beginning with under 100 confirmed cases by early March, saw the outbreak resurge with over 100 confirmed cases per day in late March and approximately 1000 per day in late April (https://github.com/stccenter/COVID-19-Data/tree/master/Global). While it was found to be transmitted especially among the disadvantaged laborers in the city state, the outbreak triggered an alarm against reopening economies around the world.
- Africa is becoming an epicenter since late April, with an estimated 120 million people predicted to be infected eventually and millions of lives to be lost [10]. The developing economic status of these countries and the lack of health facilities and ventilators, critical life-support devices for people with severe symptoms will make the situation in Africa much worse [11].
- Most recent studies argue that the virus is not likely to disappear and will persist like HIV and other viruses. Stay-at-home measures were suggested as regular practice until a vaccine is developed and can be administered broadly [12].

With the novel virus found in almost every country, infecting over three million people and taking more than two hundred thousand lives in four months (https://covid-19.stcenter.net/index.php/covid19-livemap/), this global crisis triggered a pandemic declaration by WHO on March 11, and national emergency declarations by many countries [13]. The NSF (National Science Foundation) Spatiotemporal Innovation Center brought together a taskforce of international researchers and assembled implementation strategies to rapidly respond to this crisis (https://covid-19.stcenter.net/). The goal is to support research, save lives, and protect the health of global citizens. This paper reports our understanding and perspectives of the disease outbreak from a spatiotemporal perspective.

Section 2 introduces the spatiotemporal principles and foundations supporting our analyses and understanding of the spread of COVID-19. Section 3 analyzes the spatiotemporal patterns of the spreading of the virus around the world and in the U.S., and global and U.S. policies and administrative measures using the Oxford stringency index [14]. It assesses the consequences of the outbreak to the environment, economies, human mobility patterns, and societies, and to forecasting and strategy-setting. This section reflects our efforts in dealing with a moving target, with more data emerging daily, revealing new characteristics, and patterns of the disease. Section 4 concludes with the spatiotemporal indicators, followed by reflections on the pandemic's impacts on human society and the post-pandemic world in Section 5.

## 2. The spatiotemporal matrix and spreading dynamics

### *2.1. A spatiotemporal matrix*

COVID-19 has impacts on every aspect of human society and has produced many phenomena that can be detected by different sensors, resulting in what has been termed big spatiotemporal data [15], including case numbers reported in near real time by government agencies, news agencies, and non-governmental organizations; government orders and administrative measures captured by official documents and news reports; human activities changed and reflected by social media and transportation activities; economic activities tumbling as reflected by stock markets around the globe; and the Earth-observation data reflecting a changed Earth. These datasets are big by themselves, with the five Vs intrinsically integrated (volume, variety, velocity, veracity, and value)[16]. Through an Amazon cloud credit grant, we employed the Amazon cloud to tackle these big spatiotemporal data challenges. The process can be represented by a spatiotemporal matrix (Figure 1):

- The core is the data collected about the virus, its spreading, the responses, and their impacts as variables.
- Space, in the form of place names, refers to the two or three-dimensional spatial world.
- The temporal dimension represents the fast-changing situation at the granularity of minutes, such as the changing values of the reported cases.
- The variable refers to different types of data value captured and shared (https://covid-19.stcenter.net/index.php/data-access/) including (confirmed, tested, death, recovered) case numbers, government measures, human movement, socioeconomic dynamics, air quality, logistics supply chain, available hospital beds, and many other factors.
- The core matrix supports the outer matrix belt of a variety of spatiotemporal analytics in various research fields, domains, sectors, and the society at large in the era after COVID-19.

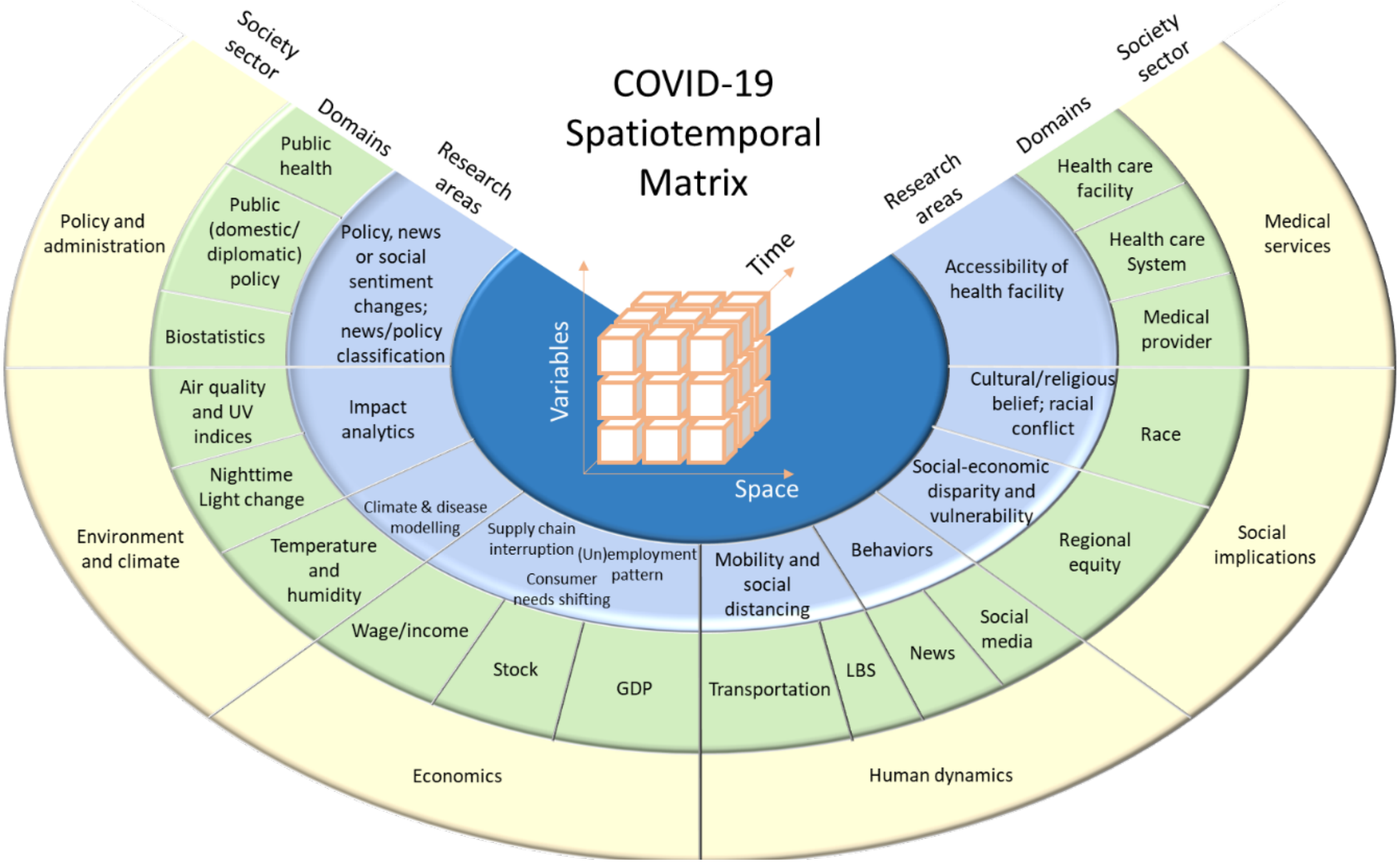

**Figure 1.** The spatiotemporal matrix integrates conceptually the space, time and variables represented by data, analytics for different research subjects, application domains, and society sectors, as human societies progress towards the era after COVID-19

The data are collected and made openly available at https://covid-19.stcenter.net/index.php/data-access/. The following sections introduce the spatiotemporal patterns identified when coupling COVID-19 data with relevant research as denoted in Figure 1.

## *2.2. The evolution of confirmed cases & worldwide spreading*

Since the outbreak of COVID-19 in China at the end of 2019 (Figure 2a), the disease spread quickly from China to other countries and territories since late January (Figure 2b) and had affected populations on every continent as of April 28, 2020. The epicenter of the outbreak shifted from China, Central, and East Asia to Europe and North America in the past three months (Figure 2c). Confirmed cases per 100,000 increased significantly in countries such as Japan and South Korean in addition to China in early February. In mid-February, the epicenter shifted to the Middle East and Europe and confirmed cases in countries such as Iran, Spain, and Italy continued to climb. The worldwide pandemic was accelerating. Since early March, confirmed cases per 100,000 increased sharply in the United States, making it a new epicenter of the crisis (Figure 2c). As of April 28, 2020, confirmed cases in five countries are more than 500 per 100,000 which includes the San Marino, Holy See, Andorra, Luxembourg and Iceland and 7 countries with over 100k confirmed cases (Figure 2d).

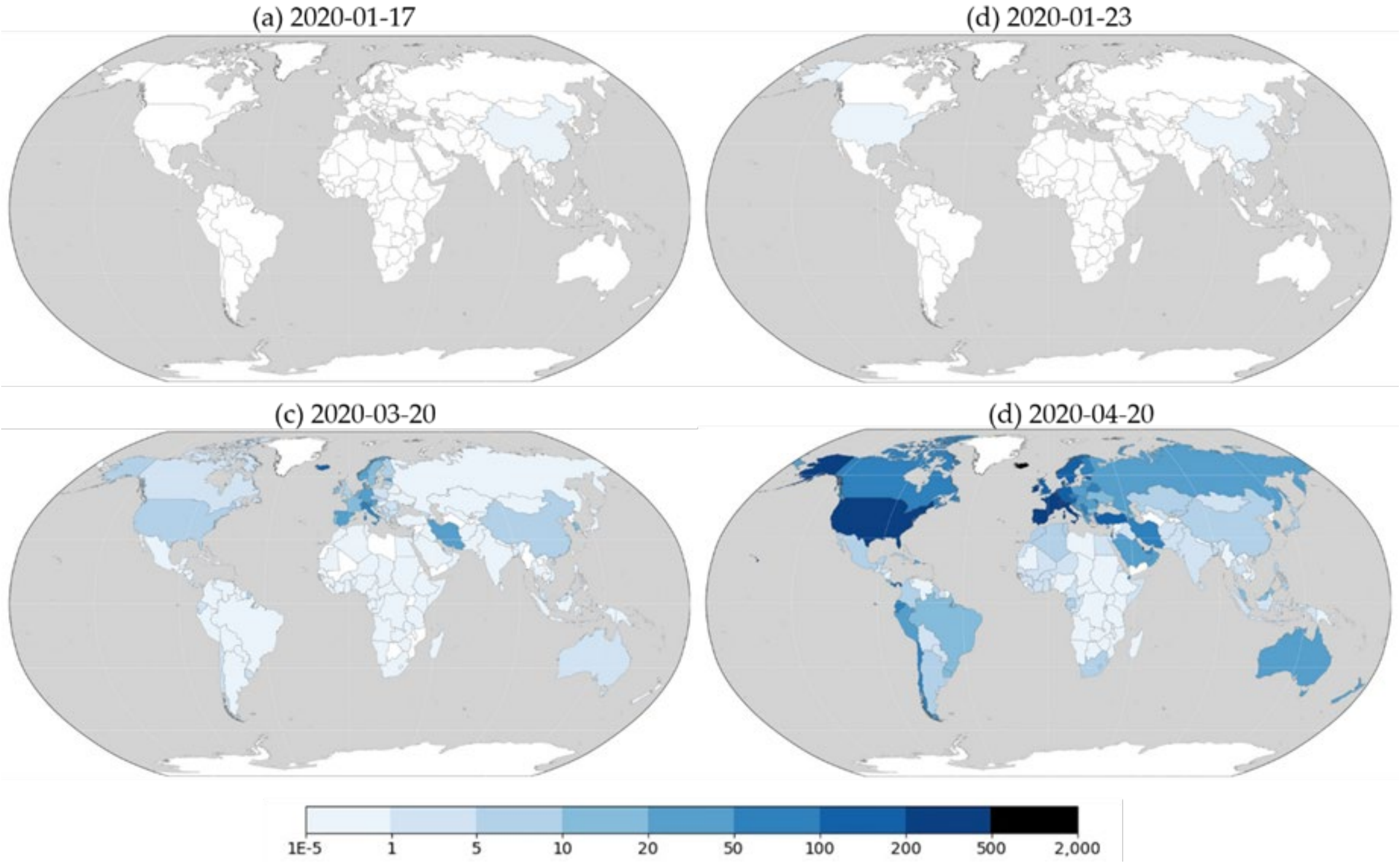

**Figure 2**. Number of confirmed cases per 100,000 across the world with animation available. (data collected from global resources, processed, and visualized by the NSF Spatiotemporal Innovation Center).

Figure 3 shows temporal snapshots of hotspots of confirmed cases in the U.S. from January to April and demonstrates the transmission of COVID-19 in the US and the shift of its epicenter. The first U.S. confirmed case was reported in Snohomish County, Washington on January 19, 2020 (Figure 3a). Since then many cases have been confirmed in different states, with New York became the world epicenter by mid-March (Figure 3c). In summary, the disease started to emerge on the east and west coast and was then transmitted to other states in the transportation hub regions of the country. In February and early March, epicenters were concentrated in cities such as Seattle, Los Angeles, and Boston (Figure 3b, 3c). Since mid-March, the outbreak of COVID-19 has spread into other cities such as New York and San Jose, and by March 17 the coronavirus was present in all states (Figure 2c). Hot spots can be found in most states in late April. As of 04/27/2020, over one million confirmed cases were reported in the U.S. (Figure 2d).

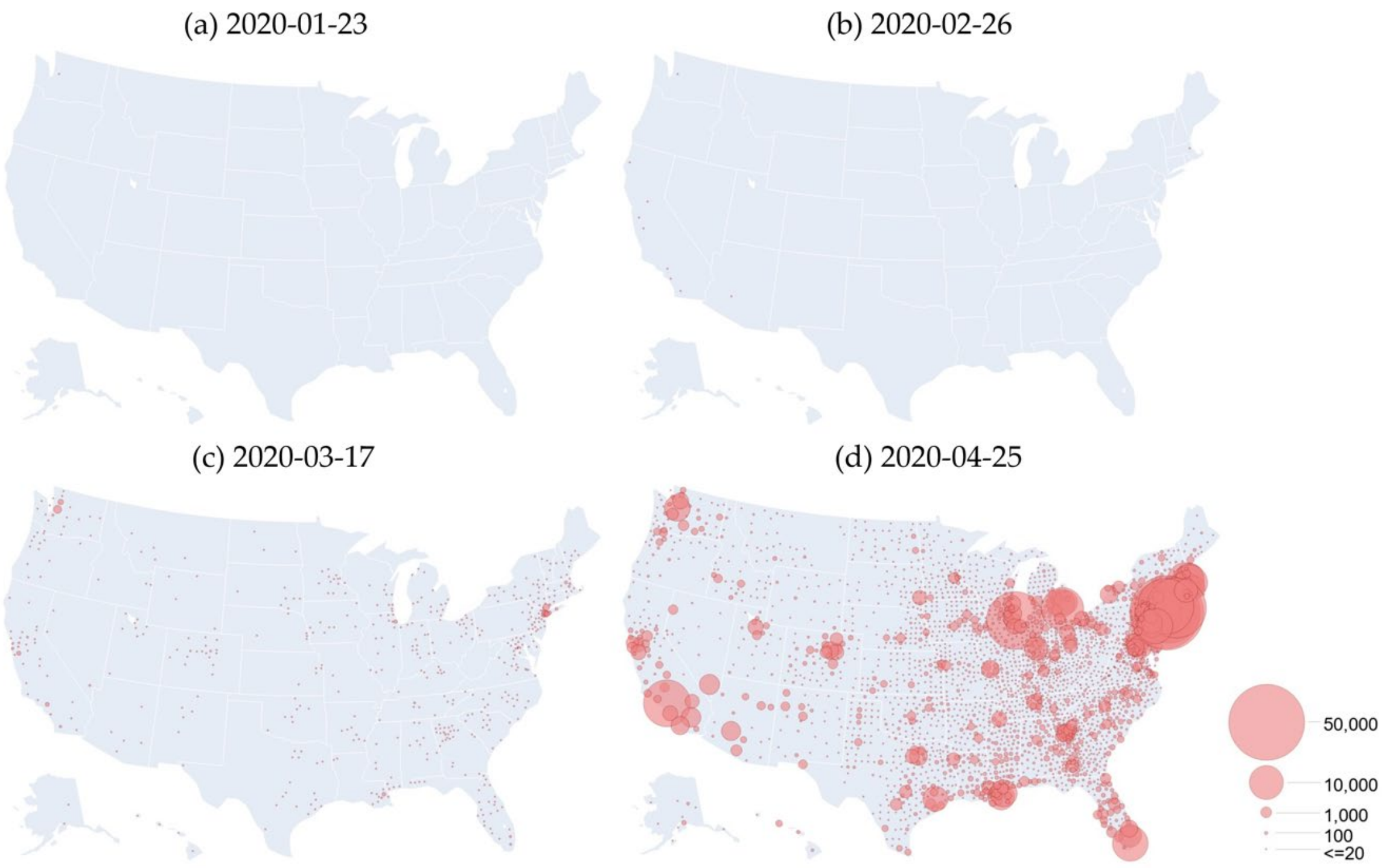

**Figure 3.** Hot spots of confirmed cases in the U.S. with daily animation available (data collected from global resources, processed, and visualized by the NSF Spatiotemporal Innovation Center).

## 3. The spatiotemporal responses

### *3.1. The responding policies and administrative measures*

The lockdown of Wuhan and China dropped the R0 from 2.x to 1.0 in February [3] and the lockdown, which may be viewed as an effective policy, eventually helped contain the outbreak in China. Border controls and lockdowns were adopted by many countries to control the spread of COVID-19. These measures need to consider that people might be asymptomatic but are already contagious during the early stages of an infection. It was also suggested to trace contacts outside of the epicenter to limit human-to-human transmission [17]. After an initial spread and subsequent containment which resulted in success in containing the virus, it is important to develop testing capabilities to detect possible resurgence of the virus. The final goal would be to develop a vaccine, which is the ultimate means to eradicate the disease. Before that, stringent stay-at-home lockdown measures have become normal in many countries to limit the spread. The strictness of relevant administrative orders measured using the stringency index becomes a criterion of the effectiveness of disease control [14].

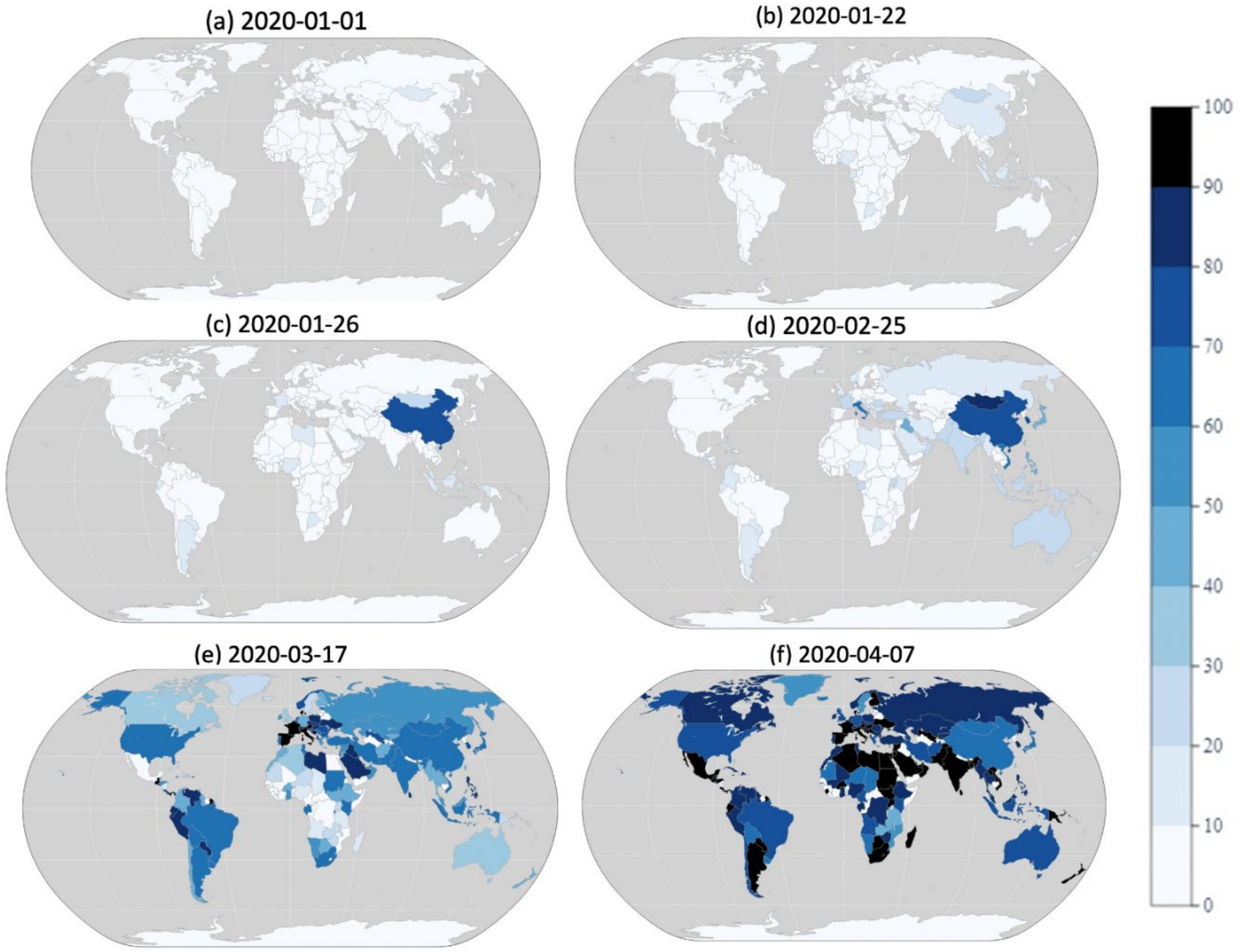

**Figure 4.** The world stringency index from January to April is increasing and correlates well with the disease spreading with a continuous daily animation. (data collected from global resources, processed, and visualized by the NSF Spatiotemporal Innovation Center).

The Government Response Stringency Index was first introduced by the Oxford COVID-19 Government Response Tracker [14]. It codes qualitative policies into numbers and then takes the average of these specific policies such as school closure, business closure, public event cancellation, as well as a generalization code to indicate the scope of the specific policy. Therefore, the index can present a good understanding of governments' responsiveness towards the current crisis. Figures 4 and 5 illustrate this index for countries around the world and the U.S. As shown in Figure 4a, the first country with a policy response was Mongolia by releasing daily news to inform the public about COVID-19. On January 22, China locked down Wuhan (Figure 4b) and issued policies included the prohibition of large gatherings and extending the Spring Festival holiday for both schools and organizations on January 26 (Figure 4c). On January 30, WHO declared a global health emergency, and by February 25, COVID-19 was detected in more than 11 Asian countries (Figure 4d). After that, more countries took action in response to WHO declaring COVID-19 "a pandemic" on March 11. As COVID-19 started to spread across the globe and declined in China by March 17, many countries took more serious measures while China relaxed its restrictions (Figure 4e). Italy and Spain announced a lockdown, Canada closed its border and the U.S. President Trump declared a national emergency. By April 7, most countries around the world had a stringency index of higher than 50, meaning they had issued harsh restrictions such as closing schools and business (Figure 4f). The effectiveness of the lockdown in China helped it reopen the economy and lifted its last city lockdown for Wuhan on April 7 after already reopened other cities in March.

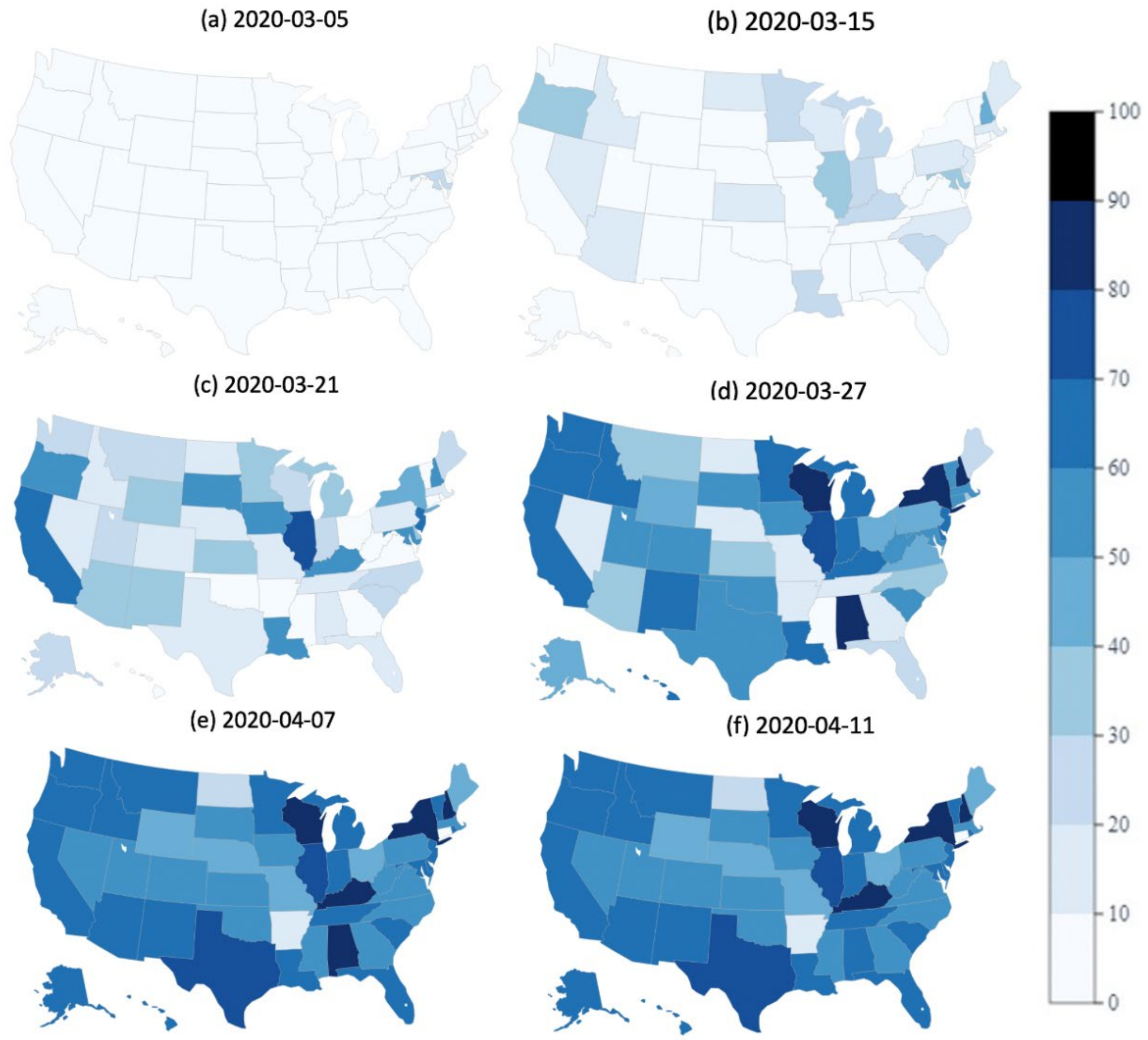

**Figure 5.** The U.S. states' stringency index increased from early March to late April and correlates well with the disease spreading; [a continuous daily animation]() is available. (data collected from global resources, processed, and visualized by the NSF Spatiotemporal Innovation Center).

The U.S. took initial action on January 31, by implementing travel restrictions from China. A further travel advisory was issued at the end of February for Italy and eventually for all of Europe. In terms of local measures, on March 5 Maryland responded to the pandemic by declaring a state emergency (Figure 5a), releasing an information campaign about COVID-19, and closing all K-12 schools. On March 15, 17 states issued orders in response to the pandemic, with western and northeastern states having the highest stringency index, since they were the states with the most cases at that time (Figure 5b). By March 21 (Figure 5c), more states issued COVID-19 related orders and the stringency index rose with positive cases in each state. As the pandemic continued to spread across the country and death tolls continued to rise, by March 27 there were 9 statewide and 10 regional "Stay at Home" orders. At that point, U.S. coronavirus cases hit 100,000 (Figure 5d). In order to stop the pandemic by reducing personal contact, 42 states had issued a statewide "Stay at Home" order and 3 states had issued a regional one by April 7, which closed nonessential businesses and prohibited large public gatherings (Figure 5e). Four days later, by April 11 (Figurer 5f), every state was under a disaster declaration simultaneously for the first time in the US history as the US death toll overtook Italy's and reached 20,000. After several weeks of shutting down, Florida was the first state to ease its restrictions and opened its beaches. Following that, more states planned to reopen their businesses in May or June.

### *3.2. The impacts*

One of the policies adopted by most government authorities to combat the spread of Covid-19 is the stay-at-home policy. Keeping the population home-bounded introduces many negative impacts, particularly along the economic front. However, keeping people at home reduces the use of automobiles, thus tremendously lowering emissions and leading to some important environmental consequences.

### 3.2.1. Human movement patterns in the shadow of COVID-19

Human movement is an important driver of the dispersion of infectious diseases [18]. The magnitude and scale of human movement are critical for the prediction of virus transmission, the identification of risk area, and decisions about control measure [19]. Different data sources, such as public transportation (bus, train, and flight), social-media data, and mobile-phone data, can be used to detect such movement.

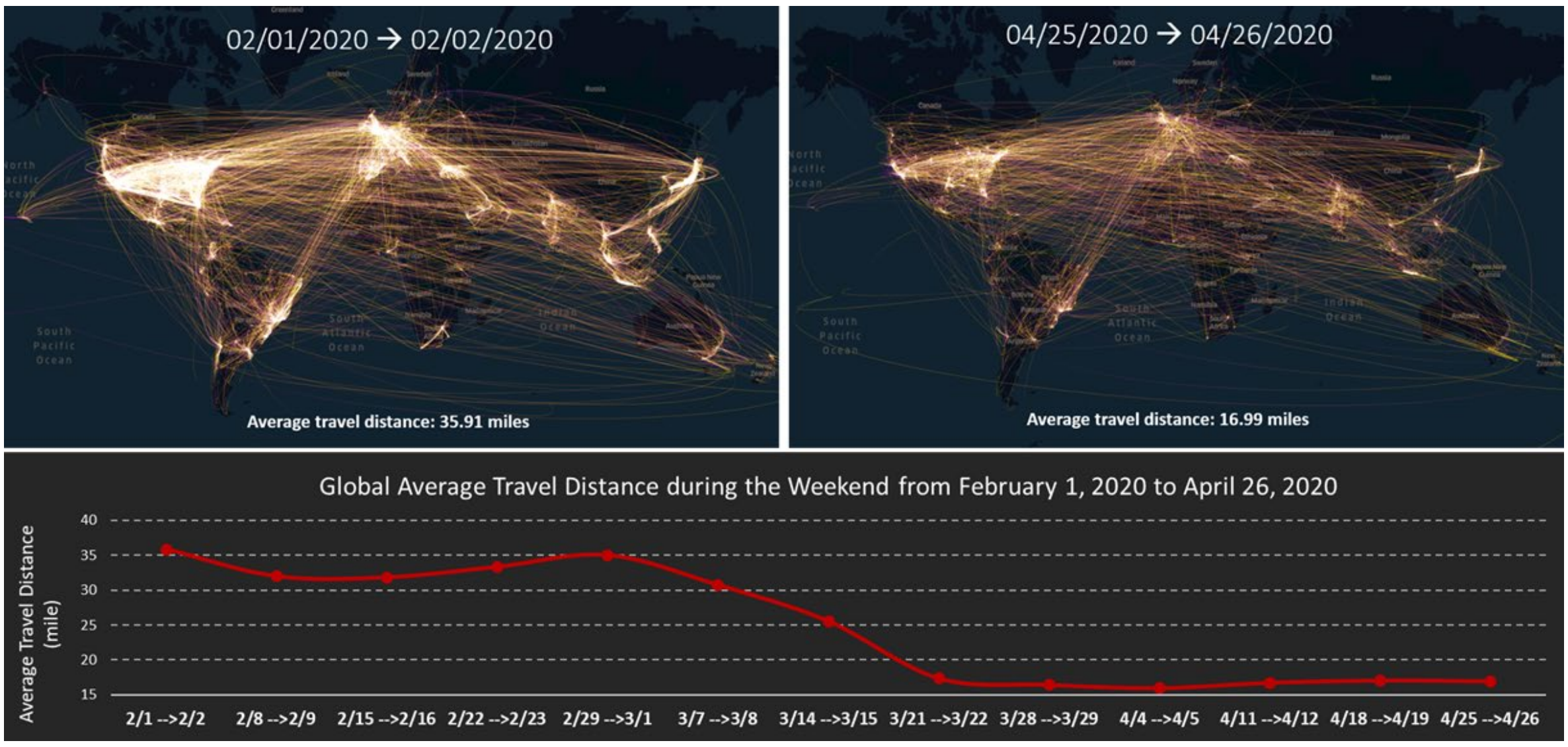

**Figure 6.** Global average travel distance during the weekend from 02/01/2020 to 04/26/2020, as derived from geotagged tweets. An animation is available. (data collected from Twitter, processed, and visualized by Z.L.)

Geotagged Twitter data have been used in human mobility studies [20-22]. Figure 6 shows the global population flows and average travel distance derived from geotagged tweets during the weekend (Saturday --> Sunday) from February 1, 2020, to April 26, 2020. It is clear from the map animation and Figure 6 that the intensity of population movements declined dramatically from February 1st (35.91 miles, Figure 6 left) to April 26th (16.99 miles, Figure 6 right), a 53 percent drop, though it started to level off after March 29th.

As the COVID-19 broke out in the U.S. in mid-March, the intensity of population movement declined dramatically from March 12 to March 24 (Figure 7). Specifically, after U.S. states issued stay-at-home orders, with California taking the lead on March 19, population flows became sparse, especially in places in the west and south coasts, such as near San Francisco, San Diego, and Miami. Also, in Texas, there is an obvious network triangle between Dallas, Houston, and Austin before March 18, and it became much weaker on March 24. However, for places near New York City, we can still see strong population movements after lockdown started on March 22. These movement changes reveal the impact of COVID-19 on people's travel. A more comprehensive analysis and modeling at finer spatiotemporal resolutions are needed to better understand how the movement patterns are associated with the spread of the virus and with the implementation of policies.

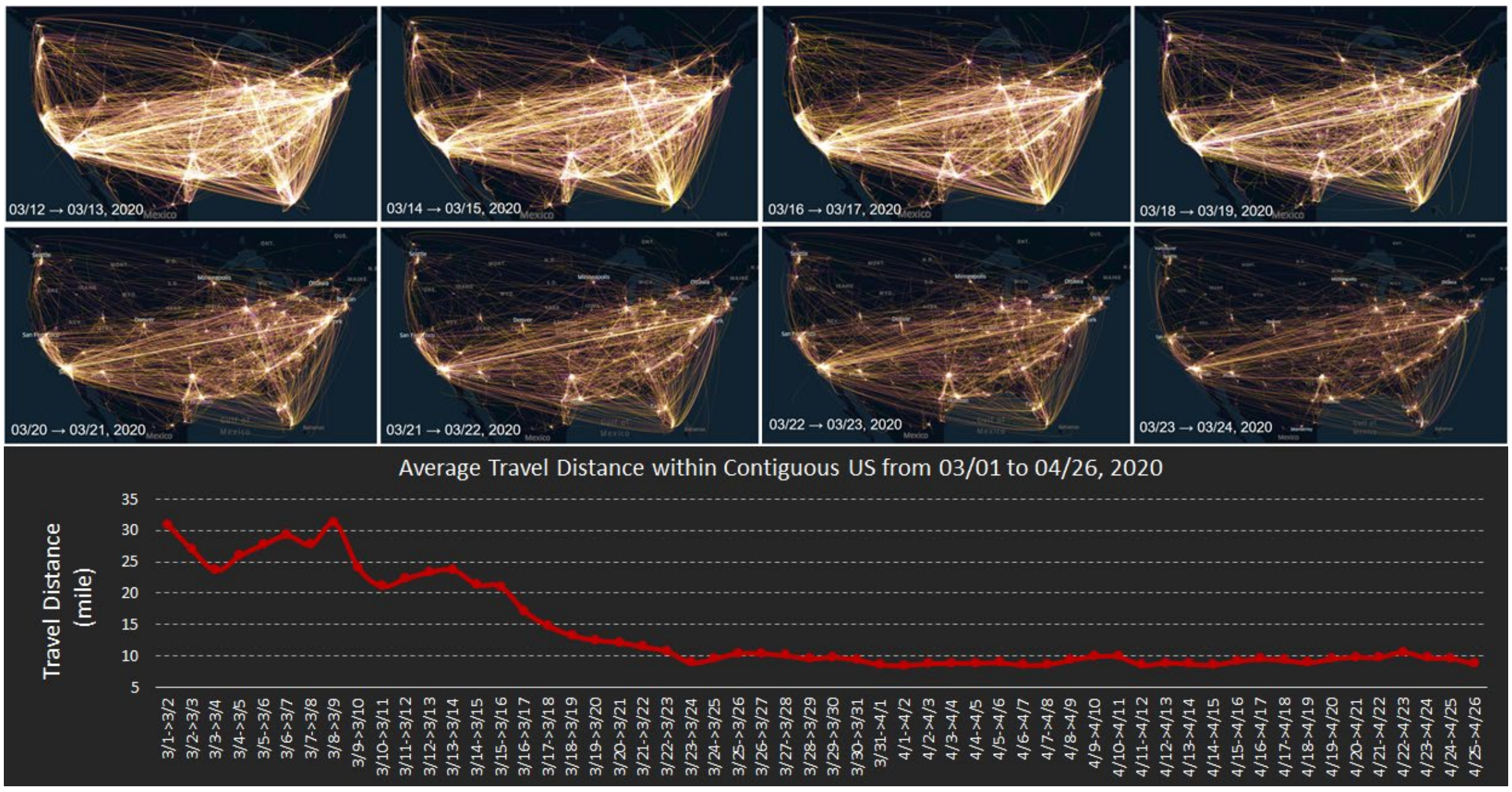

**Figure 7.** Population movements derived from geotagged tweets before and after the lockdown in the contiguous US. Brightness indicates movement intensity (more population flow in and out). Data were collected with the Twitter API. (data collected from Twitter, processed, and visualized by Z.L.)

Figure 8 shows the overall flight frequency decreasing in China, and less travel in February with many flights canceled. For example, the flight frequency on February 18 is about 31.1% lower than that of January 1. The outbreak has a significant impact on air travel, which reflects efforts in containing the virus spread. The inclusion in the analyses the number of passengers on each flight will likely reveal much less travel, as news reported that many flights were almost empty after the lockdown policy put in place.

The mobility of global communities also changed significantly as the result of a series of measures taken to combat the crisis. Aggregated service usage data from Google (https://www.google.com/covid19/mobility/) provide insights into changes in the movement trends over time in different countries across different categories of places such as groceries and pharmacy, parks, transit stations, retail and recreation, workplace and residential. Figure 9 shows that the visit to public places such as parks and groceries decreased significantly as the outbreak of COVID-19 swept across the world and social distancing policies were implemented in some countries. Activities at home increased significantly as a result of stay-at-home or work-from-home policies. The spatiotemporal patterns correlate well with the spatiotemporal dynamics of the outbreak and relevant policies.

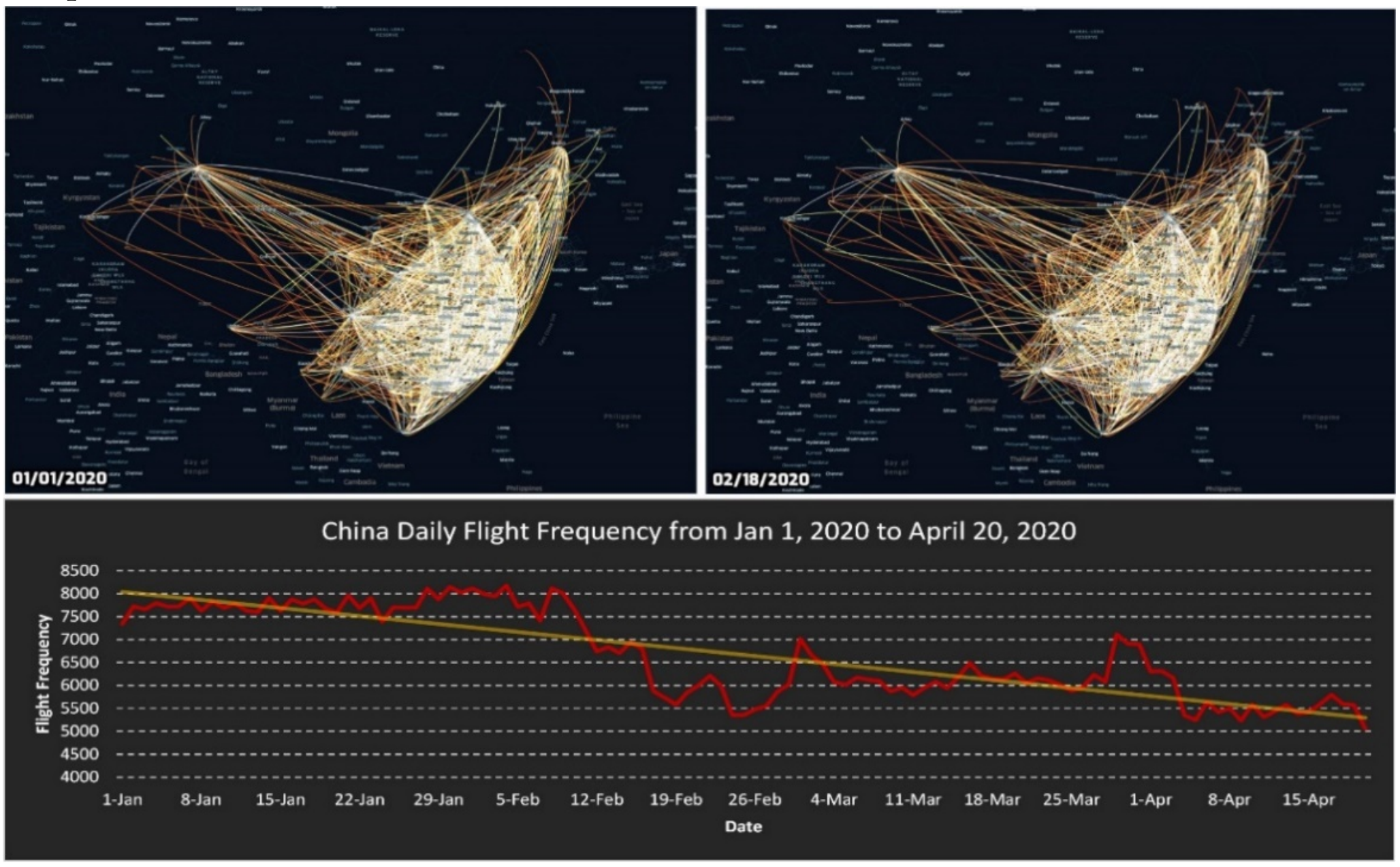

**Figure 8.** Flight frequency before and after the lockdown in many cities of China (the brightness in the network indicates the flight frequency (data collected by Wuhan Univ., processed, and visualized by the NSF Spatiotemporal Innovation Center).

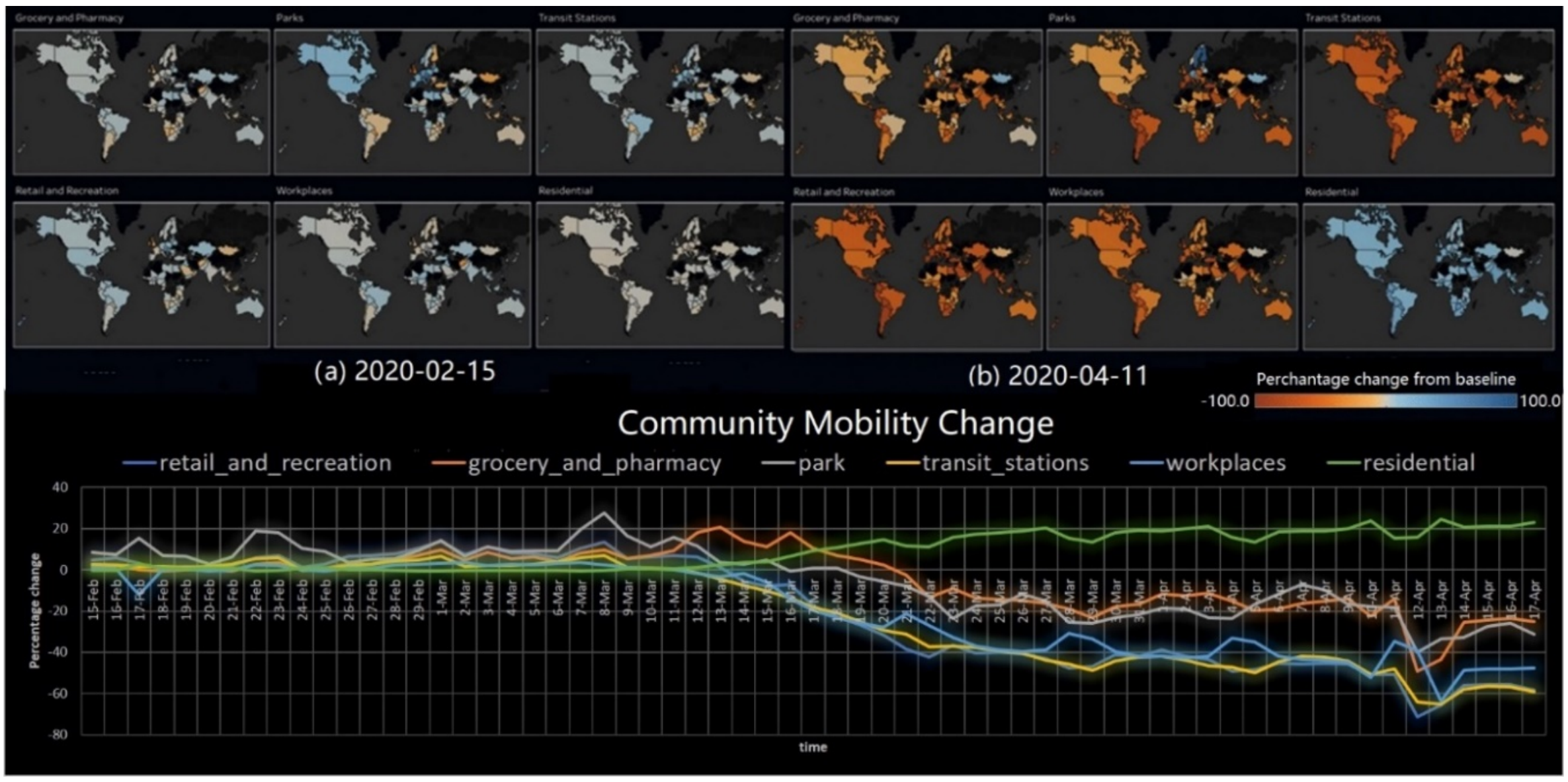

**Figure 9.** Percent change of daily community mobility with [animation]() available. (data provided by google, processed, and visualized by the NSF Spatiotemporal Innovation Center)

### 3.2.2. A quieter and cleaner Earth

Due to the lockdown of many locations, and thus tremendously cut back automobile usage and shut down factory operations, the amounts of various types of pollutants released to the environment were significantly reduced. As a result, air quality has significantly improved since the COVID-19 outbreak [23]. The concentration of nitrogen dioxide and air quality index (AQI) are analyzed using ground-based observations. Figure 10 shows the hotspot maps based on kernel density estimation of nitrogen dioxide concentration which is the primary pollutant emitted by motor vehicles, power plants, and industrial facilities, and the mapping bandwidth is 80 kilometers. It diminished in density over China when most cities shut down during February compared to January and late March of 2020.

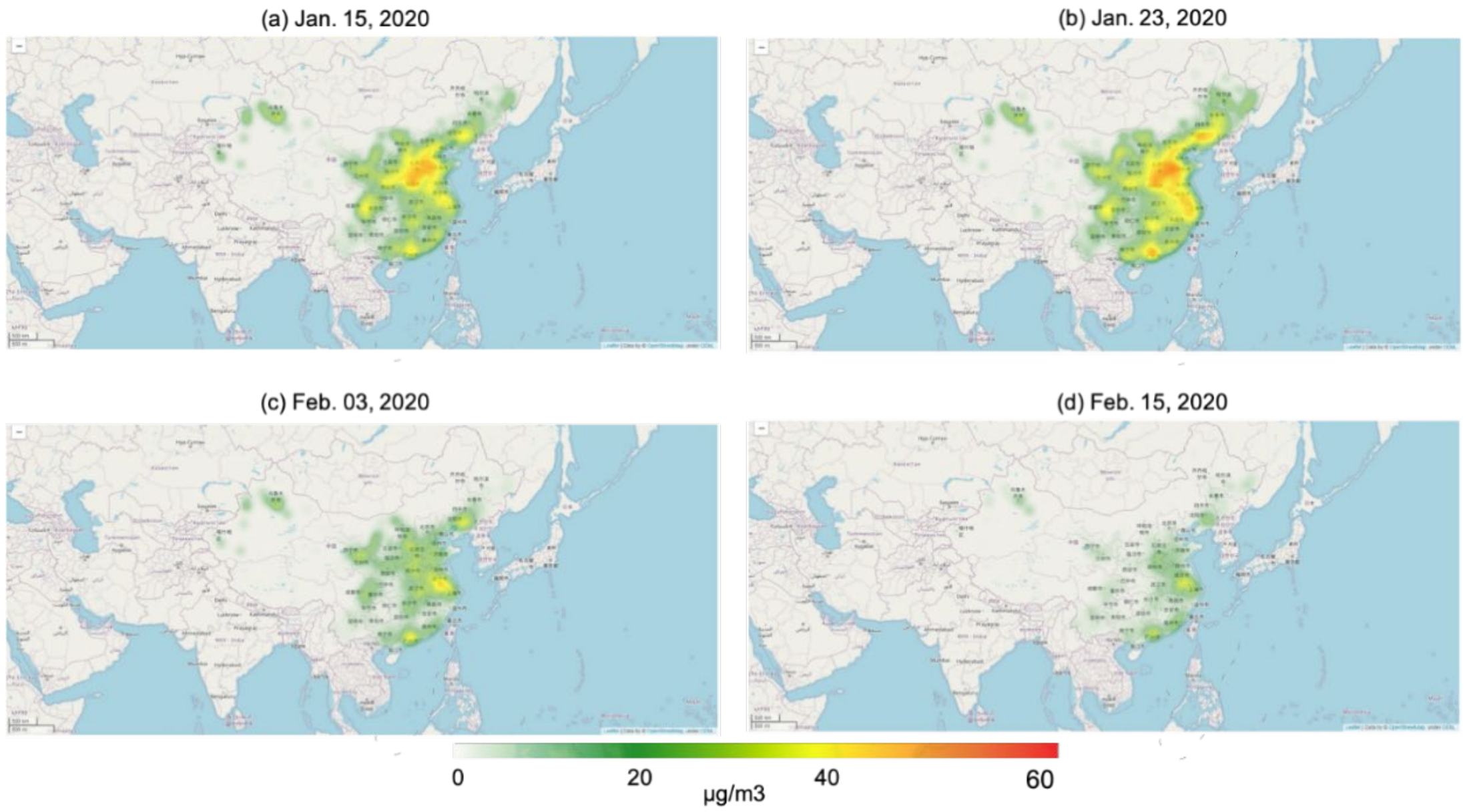

**Figure 10.** Nitrogen dioxide concentration before and during the pandemic in China with daily animation (ground-based data collected from China National Environmental Monitoring Center, CNEMC, analyzed and visualized by NSF spatiotemporal innovation center).

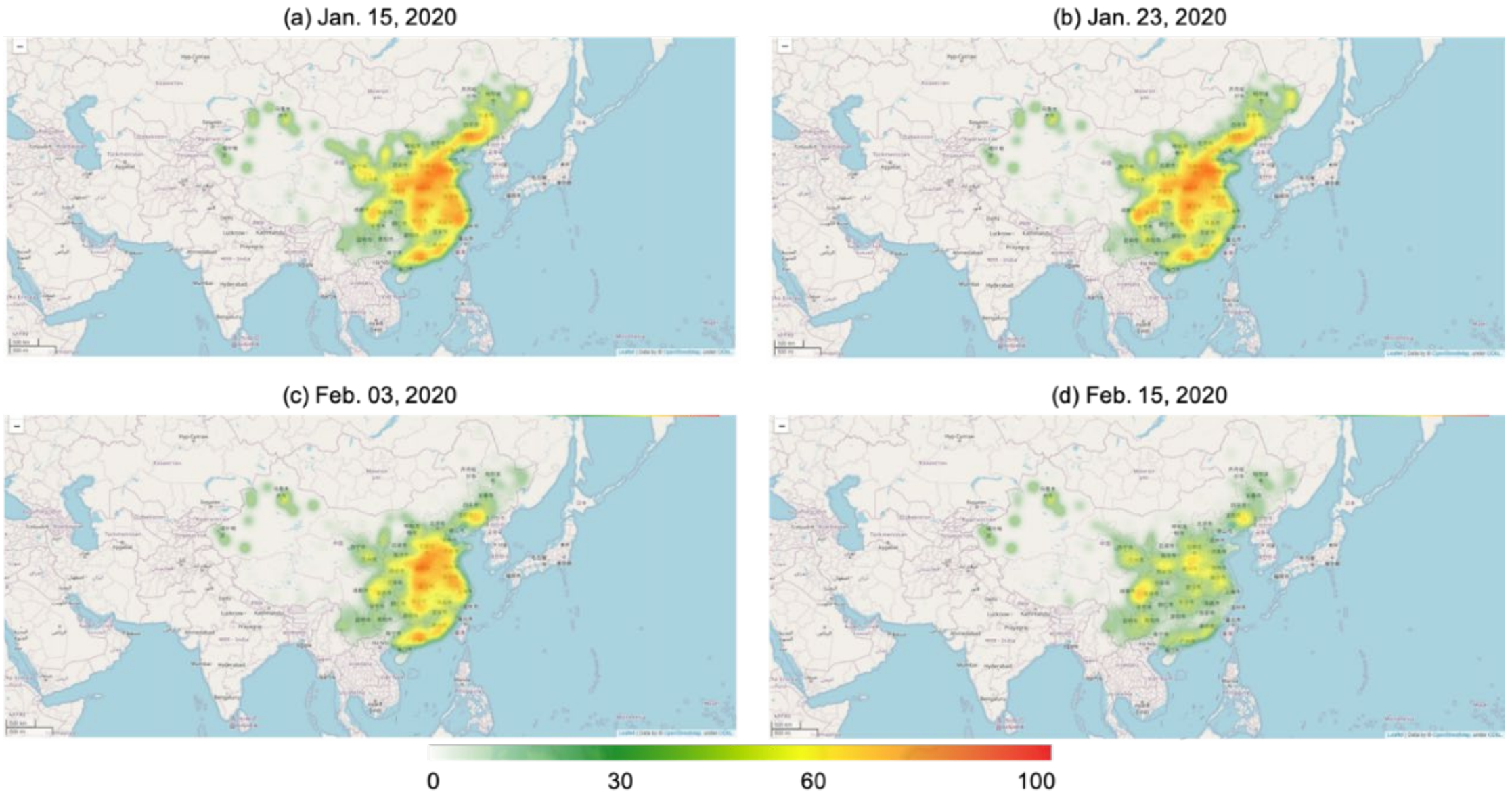

**Figure 11.** AQI before and during the pandemic in China with daily animation (ground-based data collected from China National Environmental Monitoring Center, CNEMC, analyzed and visualized by NSF spatiotemporal innovation center).

Similar phenomena are also observed with an air quality index (AQI) (Figure 11), which records the intensity of air pollution in China by considering five main air pollutants: PM2.5, PM10, Ozone, Nitrogen Dioxide, Sulfur Dioxide and Carbon Monoxide [24]. According to the U.S. Environmental Protection Agency (EPA, https://www3.epa.gov/ttn/oarpg/t1/memoranda/rg701.pdf), AQI is the highest index of all the six types of pollutants using the following equation:

$$\text{AQI} = \text{Max}\ (I_i),\ \{i = 1,2,3,4,5,6\}$$

$$I_i = \left[\frac{I_{i,hi} - I_{i,low}}{BP_{i,hi} - BP_{i,low}}\right](C_{i,P} - BP_{i,low}) + I_{i,low}$$

where $I_i$ is the index of one type of pollutant ; $C_{i,P}$ is the rounded concentration of pollutant i; $BP_{i,hi}$ is the breakpoint greater or equal to $C_{i,P}$; $BP_{i,low}$ is the breakpoint less than or equal to $C_{i,P}$; $I_{i,hi}$ is the AQI corresponding to $BP_{i,hi}$; $I_{i,low}$ is the AQI corresponding to $BP_{i,low}$.

The AQI values decreased during February of 2020. The reduction of air pollution was first shown near Wuhan and then extended to most parts of China. These changes were due to the quarantine and shutdown policies which limited the usage of motor vehicles and industrial productions.

Figure 12 shows the Night-Time Light (NTL) changes of Hubei Province before (a) and during (b) the lockdown. Commercial and business centers are shown to be dimmer in February than in January because of the spreading of the virus and the social distancing policies. People stopped working, getting together or congregating outside, and instead stayed at home. Light pollution is reduced during the COVID-19 crisis. Light pollution from some of the highways between cities is less intense during the pandemic due to the shutdown of cities and cut off of transportation.

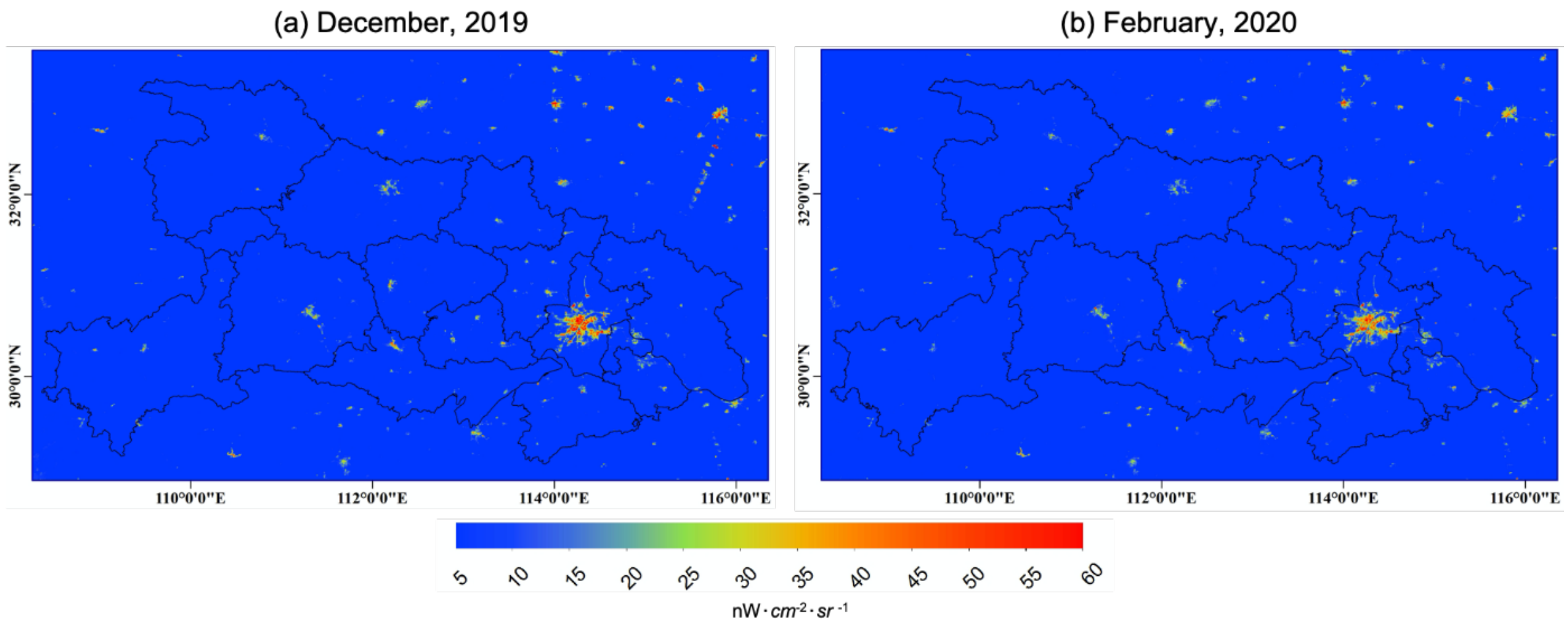

**Figure 12.** Nighttime light of Hubei Province before and during the pandemic with daily animation (data collected from NASA's LAADS DAACs, processed, and visualized by the NSF Spatiotemporal Innovation Center).

3.2.3. The economic impacts

The global economy and stock market have also been significantly impacted. Stock markets worldwide performed according to the timeline of the virus outbreak in each region (Figure 13). Nine representative stock indices were selected to cover Asia's, Europe's, and America's major stock markets (Figure 13). 25 stocks were chosen to represent the global stocks traded in New York Stock Exchange. From January to April, the global market responded dynamically with the global transmission of the COVID-19. When the first cases were reported in China and Wuhan was locked down, the Chinese Mainland SHA, SS, and HSI indices dropped significantly with SS plunged and SHA fluctuated during the entire observed period. The global market showed a slight decrease before mid-February. The Chinese stocks (such as Alibaba and China Netease in Figure 14) traded in NYSE showed similar performance like HSI. After China announced that the outbreak was under control in February, the SS and HIS showed an upwards trajectory. The global market dropped in late February and early March with the European indices (FCHI, GDAXI, and FISE) leading the trend when Europe became the epicenter. All indices hit the bottom and started gaining in late-March after a global announcement of stimulus packages for COVID-19, following the U.S.

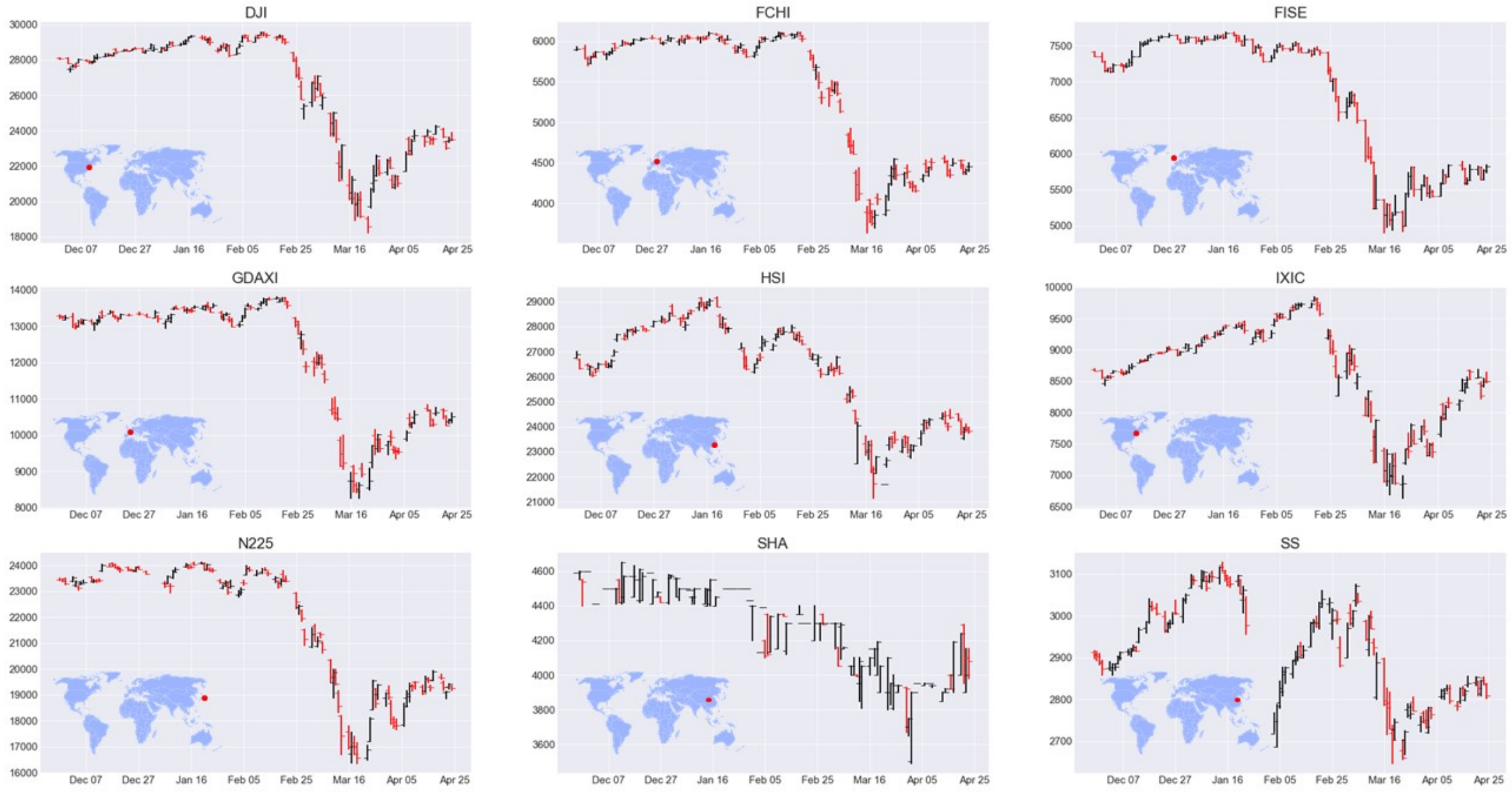

**Figure 13.** The global market has various spatiotemporal patterns corresponding to the spreading of the virus (data collected from global stock markets, processed, and visualized by the NSF Spatiotemporal Innovation Center

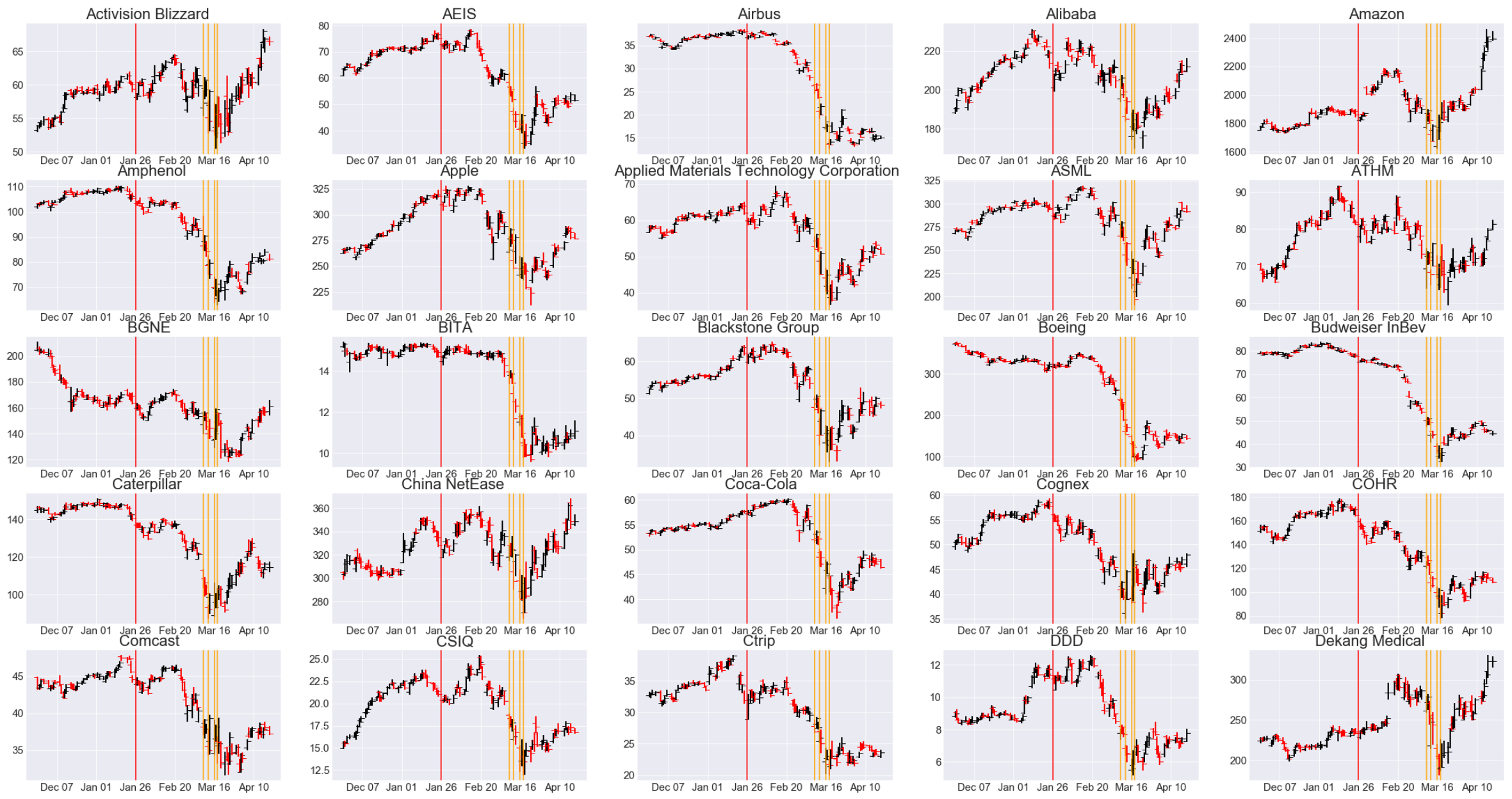

**Figure 14.** Behavior of representative stocks' on the U.S. market. The red vertical line shows when the first case was confirmed, and four trading curbs that happened in early March were added as yellow (data collected from New York Stock Exchange, processed, and visualized by the NSF Spatiotemporal Innovation Center)

American markets (DJI, IXIC) stopped rising in mid-February and plunged until mid-March. All representative stocks for different industry sectors, such as high-tech, aviation, medical, and food, showed a slight drop when the first case was reported in the U.S., and plunged in March. While most stocks did not respond much to the late January and February lockdown of China, Chinese stocks (such as Alibaba and Ctrip) plunged for the first time in 2020. April saw a gradual increase of the market with some sectors (e.g., Activision Blizzard, Amazon, and Medical for health and logistics) quickly recovered, while some (e.g., Airbus and Ctrip for travels) were still at the bottom, based on each industry's market resilience (Figure 14).

The global market performance demonstrated a tightly coupled world village but also regional differences and the sensitivity of the market to regional and global events. This correlates well with the GDP announced later. For example, the Chinese GDP dropped 30% in February 2020 and Hubei GDP in February 2020 dropped 98%. The 2020 1$^{st}$ Quarter GDP dropped 4.8% in the US.

3.2.4. Social implications

The coronavirus does not discriminate; however, disadvantaged and vulnerable social groups are bearing a larger brunt of the infection and death toll [25]. China CDC reported that 89.8% of the confirmed cases in Wuhan were between the ages of 30-79, while 88.6% of those in Hubei and 86.6% of those in China before February 11, 2020 [26]. For mortality, most of the cases in Italy are in the age group of 60+, and more than 50% of mortality cases in China are in the age group of 50+. Thus, older people have a higher risk of being infected and succumbing. In addition, different races have different risk levels and factors [27]. By April 23, 2020, African Americans' COVID-19 mortality rate was 2.7 times higher than that of Whites, 2.5 times higher than that of Asians, and 2.4 times higher than that of Latinos. (https://www.apmresearchlab.org/covid/deaths-by-race). At the state level in the U.S., the death rate (based on weighted population distribution) shows different spatiotemporal patterns by race and Hispanic origin (https://www.cdc.gov/nchs/nvss/vsrr/covid_weekly/). The ratio shows in pie chart (Figure 15) is the standardized death number by race population per capita, and the labelled number in center of circle shows the total accumulated death cases in each state by April 28, 2020. African Americans take the highest risk in Eastern and Southern United States and Rustbelt states, like Louisiana, Georgia, Indiana, Michigan, Missouri, Alabama and Mississippi, while American Indians have a relatively higher risk in Arizona (Figure 15). Hispanic or Latino take the highest risk in Northeast and Southern part of the county, such as New York, New Jersey, Florida and Texas. In western states, the race and Hispanic origin shows an average pattern, like California, Colorado, Nevada and Washington.

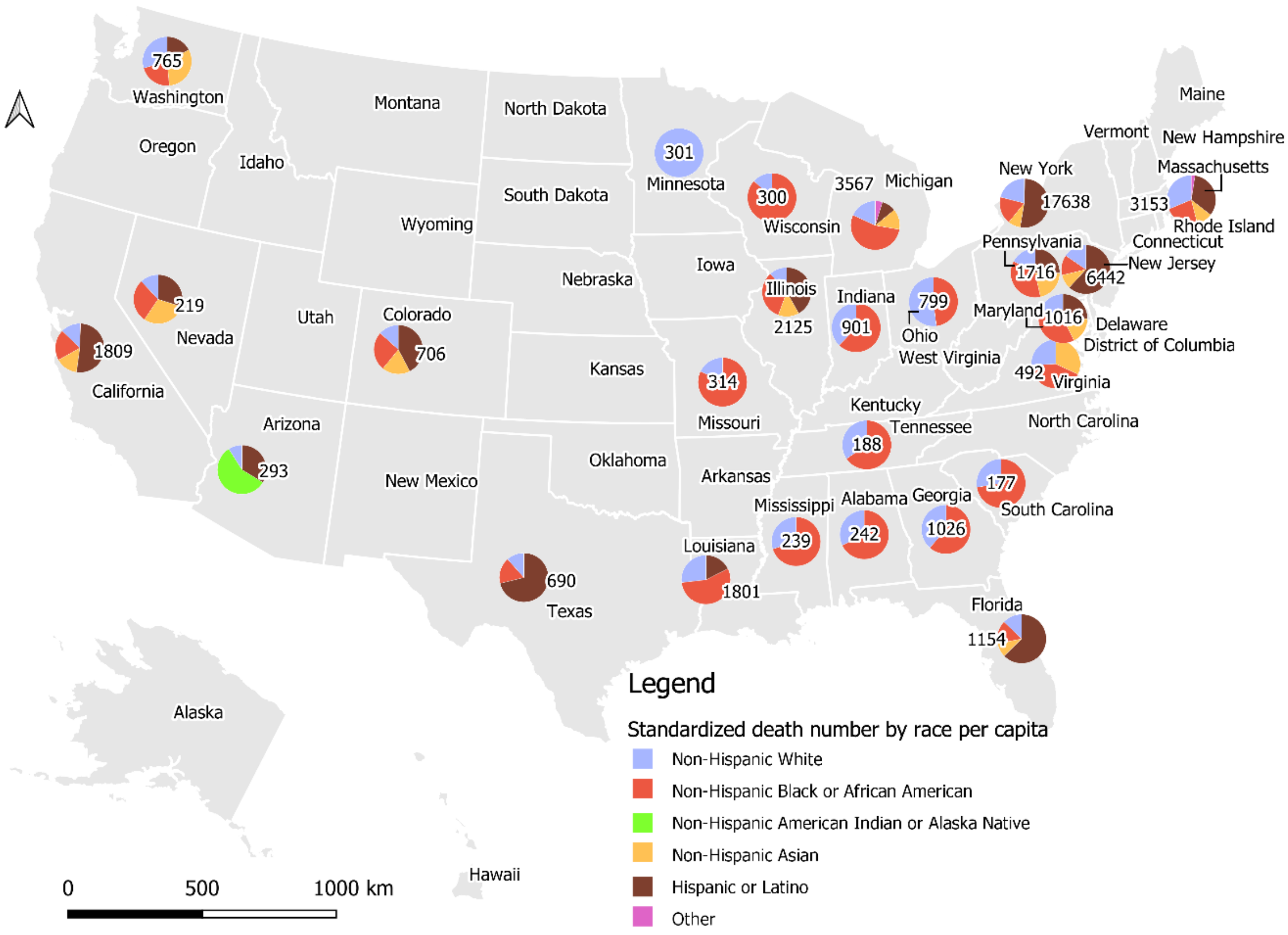

**Figure 15.** Some states with more than 100 deaths as of 4/28/2020 have strong differences by race (data provided by National Center for Health Statistics, NCHS, processed and visualized by the NSF Spatiotemporal Innovation Center).

The socioeconomic patterns are analyzed at a finer scale using weekly confirmed cases in Massachusetts by city/town (https://www.mass.gov/info-details/covid-19-response-reporting#covid-19-cases-by-city/town-) on April 29, 2020 including 350 cities/towns and 42 of them have uncertain numbers. Four variables, including poverty rate, educational attainment, elderly people rate, and income, are used to analyze the socioeconomic impact. Poverty rate is measured by the percentage of population whose income is below the poverty level; educational attainment is measured by the percentage of population over the age of 25 with Bachelor's degree; elderly people rate is measured by the percentage of population over the age of 60; and income is measured by household total annual income per 10, 000 dollars. 2017 socioeconomic data are obtained from the U.S. Bureau of the Census' American Community Survey (ACS) at county subdivision level. We employed Spatial Lag (SL) model [28] to assess the relationship between confirmed cases rates and each of the explanatory variables with the following regression equation

$$y_i = \beta_0 + \beta_1 POV_i + \beta_2 EDU_i + \beta_3 ELD_i + \beta_4 INC_i + \rho w_i \cdot y_i + \epsilon_i$$

where $y_i$ is the COVID-19 confirmed cases rate for city/town $i$, $POV$ is the poverty rate, $EDU$ is educational attainment, $ELD$ is the elderly rate, $w_i \cdot y_i$ is the spatial lag, which is calculated by weighted average of dependent variable $y$ for city/town $i$, $\epsilon$ is the random error term, and $\beta$ and $\rho$ are the coefficients to be estimated.

Table 1 presents statistical results from SL model. The R-squared is 0.577802 and the $p$ values are much smaller than or around 5%, at an acceptable level. The coefficient values for elderly rate and educational attainment are -12.6535 and -9.96245, indicating these two variables have significant and negative impact on the confirmed cases rate, i.e., well-educated and elderly population have less confirmed cases. Not surprisingly, the coefficient for poverty rate is 17.0477, representing it has a significant and positive relationship with confirmed cases rate. However, household total income has weak correlation with confirmed cases rate. In addition, spatial lag term is significant in the model, suggesting that the confirmed cases rate is dependent not only on the explanatory variables within the city/town but also on the values in adjacent city/town. Based on this preliminary study, more spatiotemporal correlation models can be involved and compared to accommodate spatiotemporal variations, such as GWR (Geographically Weighted Regression) [29] and GTWR (Geographical and Temporal Weighted Regression) [30].

**Table 1.** Spatial Lag Model-based correlation analysis results of confirmed cases rate and socio-economic variables

| Variable | Coefficient | Std. Error | Probability |
|---|---|---|---|
| Elderly Rate | -12.6535 | 5.29928 | 0.02364 |
| Educational Attainment | -9.96245 | 5.2992 | 0.05011 |
| Poverty Rate | 17.0477 | 5.99688 | 0.00447 |
| Income | 0.3127 | 0.137774 | 0.02320 |
| Spatial Lag | 0.591188 | 0.0487105 | 0.00000 |
| Constant | 4.13026 | 5.59176 | 0.00233 |
| R-squared | | 0.577802 | |

*3.3. The forecasting and strategy battleground*

With the spatiotemporal patterns detected and analyzed for COVID-19 cases and relevant policies and consequences, it is a grand challenge to forecast the future trajectories of the outbreak, simulate potential policy scenarios, and predict the consequences of reopening the economy. Fortunately, several technologies have been developed in the past decades to simulate person-to-person disease transmission, outbreaks, and relevant socioeconomic impacts by considering the transmissibility among objects with multiple factors.

Agent-based models (ABMs) are used for complex systems simulation considering individual objects as agents [31]. Individual autonomous agents, the basic elements to form the ABMs, make decisions under pre-configured rulesets during a simulation process to control individual behaviors, communication among agents, and interaction between the agents and the environment [32]. Thus, dynamic patterns could be represented by observing the ABMs as whole systems. ABMs could be applied as standalone simulators or be integrated with models in related disciplines to help strengthen existing studies [33]. For example, in the field of infectious disease epidemiology, ABM played an important role in monitoring and analyzing infectious processes [34], evaluating response policies [35], and supporting the development of containment strategies [36]. During the pandemic of COVID-19, Stevens utilized an ABM-based simulator to evaluate if containment strategies could help to "flatten the curve" [37] with four control-policy scenarios, and found that policies like extensive distancing and stay-at-home orders could help significantly flatten the infection curve.

We slightly modified the epiDEM model [38], a basic ABM-SIR model powered by NetLogo [39], to represent a similar simulation (Figure 17). This model includes three variables: S, I, and R, representing the number of susceptible people, infected people, and recovered people, respectively at specific timesteps.

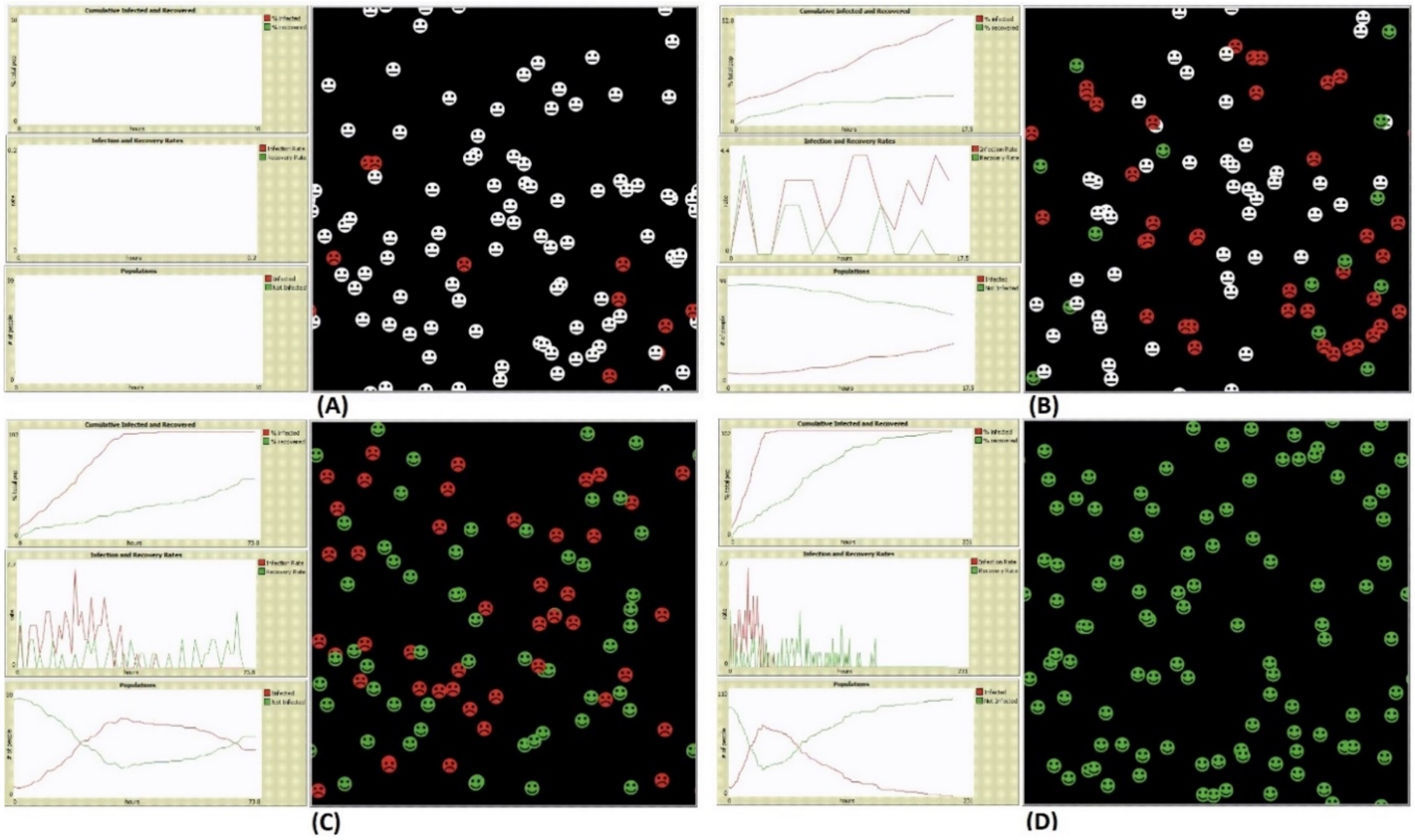

**Figure 17.** Initial ABM-SIR model simulation; an animation is available.

In addition to considering epidemiological models and policy factors, there are many other types of pandemic-effective factors, such as public-health factors (including the social and economic environment, human characteristics and behaviors, social activities, and transportation patterns)[40] and environmental factors (including temperature, humidity, and air quality)[41,42].

In previous epidemiological studies, ABMs have been used to simulate and predict the effectiveness of containment strategies under different policies [36], the time and space of outbreaks [43], medical resource deficiencies (Nap) [44] and impacts on logistics systems [45]. However, studies applying ABM coupled with multivariate impact factors, such as spatiotemporal distributions of viruses, human migration and activities, climate conditions and environmental factors, and containment strategies and policies, to reveal and predict the pandemic pattern of COVID-19 have not been developed. These factors are critical for setting up the behaviors and attributes of each individual agent and a variety of complex dynamic environments. Such criteria should consider the transmissibility (such as R0), the agent contact mode, temperature, humidity, air quality, night light, and UVs, as shown in Figure 18:

- The temperature and humidity have a strong correlation with outbreak (Figure 18a).
- The absolute humidity has a close correlation with R0 (Figure 18c).
- The night pollution has a strong correlation with economical GDP (Figure 18b).
- The outbreak spreading has a strong response to policy (Figure 18d)

However, more accurate quantitative studies are needed to mine the relationships for feeding into the overall ABM.

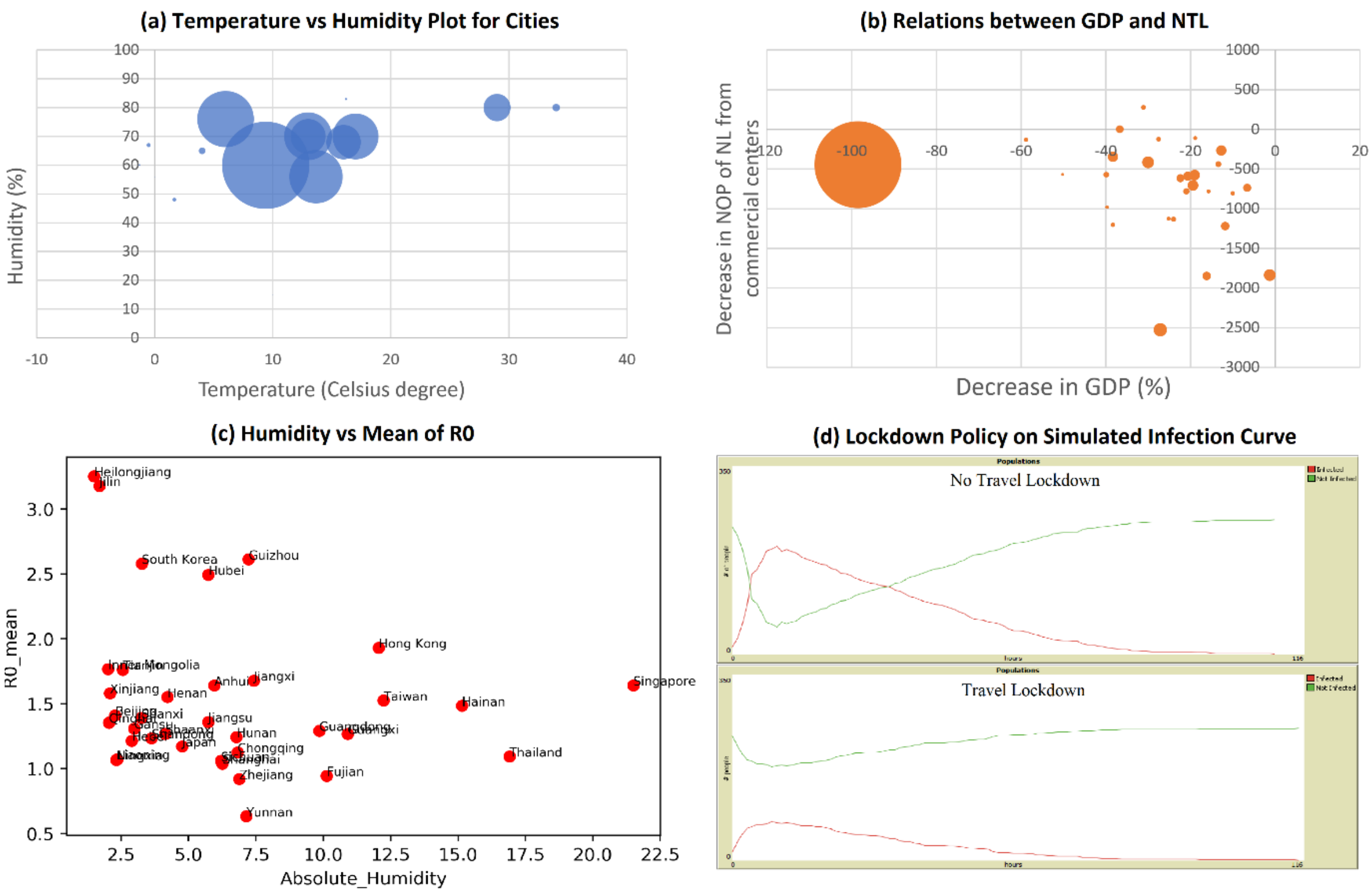

**Figure 18.** Potential relationships between the outbreak of COVID-19 and (**a**) temperature and humidity, (**b**) night-time light and the economy, (**c**) humidity and mean of R0 [42], and (**d**) control policies [38]

Statistical methods (e.g., linear regression and feature selection) can discover the relationships between impact factors (e.g., environmental factors, social-economic factors) and the outbreak of COVID-19; meanwhile, data mining methods (e.g. clustering and anomaly detection) can be utilized to mine hidden patterns (e.g., older people are more susceptible to infection; people with blood type O are less susceptible to infection) from COVID-19 related data. The learned knowledge will serve as valuable input to the ABM for simulation optimization. Figure 18 shows our initial attempts to quantitatively measure potential factors, i.e., relative humidity, average temperature, and lockdown policy, using statistical methods. Take one finding, the relative humidity (RH) was largely in the range of 45-85% in affected areas, for example, relative humidity can be set into this range in the ABM for simulation of COVID-19. We are conducting quantitative analytics to extract the relationships between COVID-19 outbreak and its socioeconomic impacts and a variety of relevant factors as a fashion appropriate to big spatiotemporal data event extraction fashion (Figure 18)[46], such as economic conditions, temperature and humidity, air quality and UV index, etc., using machine learning, clustering and regression, feature selection, anomaly detection, as well as nonlinear extensions via techniques such as generalized additive models and deep-learning methods.

Currently, we are developing and integrating a comprehensive ABM-based simulator to replicate the COVID-19 outbreak at different scales, such as the county level, state level, nation, and even global levels. We plan to calibrate our model and build region-based transition rules with multiple potential impact factors and existing COVID-19 records, by mining the possible correlation between them to find out why the COVID-19 outbreak was particularly severe in some places such as Wuhan, New York City, and northern Italy. An accuracy assessment will be conducted to verify the simulation confidence by comparing the simulated results with COVID-19 records. We can then apply our simulation model to describe the multi-scale COVID-19 outbreak patterns with certain confidence levels as well as to predict the possibility of outbreaks in some places like India and South Africa, which may help mitigate the pandemic and save lives in those areas not yet hit by an outbreak, or a resurgence in reopened areas. The model will also be expanded to produce other impact factors such as economic output (18b).

## 4. Conclusion: A collective spatiotemporal perspective

The analytics reported in previous sections indicate the spatiotemporal patterns of a variety of variables:

- Disease transmission patterns: The disease transmission is fast (Section 2.2) with the epicenter moving from Wuhan to Iran, Italy/Europe in late February, and New York/U.S. in late March. The early policy and administrative orders put in place played a big role in containing the outbreak (Section 3.1). For example, China quickly controlled the spreading and reopened the economy and Singapore contained the spreading at the beginning. The opening of the economy also brought some resurgence from place to place. For example, the outbreak in Suifenhe, a small border city in northeastern China next to Russia, with 80 confirmed cases and 124 asymptomatic carriers in mid-April, demonstrated the uncertainties and difficulties in containing the disease.
- Social-economic disparity and vulnerability: While the outbreak reached almost everywhere human movement occurs, its severity and human vulnerability are quite different among different communities with different socioeconomic backgrounds, e.g., countries with stricter rules would find it easier to slow the outbreak while countries with more freedom in principle find it harder to impose strict lockdown orders (Section 3.2.4). This even triggered a worry that the virus may undermine freedom and democracy [47]. People with less access to information and with lower income are more vulnerable because of the lack of information and knowledge, difficulty in putting orders to actions, lack of means to maintain basic livelihood without going to work, and lack of preparedness in the community to respond to the outbreak.
- Mobility and social distancing: Social distancing is important and stay at home orders are key to controlling the outbreak (Section 3.1). The worldwide collective efforts of limiting travels are reflected in the maps of intensive population movement and average travel distance (Section 3.2.3). Most early-outbreak states in the U.S. with strict social distancing in place are subsequently seeing a drop in confirmed cases on a daily basis, while many southeastern states with a big number of confirmed cases could control the spreading if they applied strict rules of social distancing and lockdowns. This is also confirmed in the correlation between outbreak and human mobility in relation to the stringency index (Section 3.1). CDC reported that the selective border control did not work well because many cases were transmitted to the U.S. from Europe and other Asian countries while the transmission from China was controlled by stopping over 86% of the flights (https://www.businessinsider.com/cdc-official-says-us-missed-chances-to-stop-the-coronavirus-2020-5).
- The availability of healthcare facilities, such as hospital ICUs and ventilators, is also very dynamic. The outbreak in New York was predicted early to be needing more facilities, but later turned out to be sufficient, while many southern states without any obvious need for equipment early on turned out to be needing more later. This is based on the economic development status of specific regions, policies in place, and the absolute number of facilities. More quantitative estimation of the demand for and allocation of health facilities, supply-demand prediction, and optimization would be very helpful. Especially, more in-depth research is much needed at the current stage with an interdisciplinary approach in a spatiotemporal framework (Section 3 and especially Section 3.2.4).
- Climate has some relationship to the disease outbreak as witnessed by the similar climate zone of the outbreak cities and regions from Wuhan, Iran, and Italy to Seattle and New York (Section 3.2). This is confirmed through early studies and leads to the belief that the increase of temperature and humidity would help slow down the transmission [48] while other scholars argue that this may not be true (Section 3.3)[49].

- Economic issues, such as shifting consumer needs, interruption of supply chain, (un)employment patterns, etc., are greatly impacted by the outbreak as evidenced by the sharp stock market drop (Section 3.2.2), the national unemployment rate increase, and the almost stopped GDP in Hubei Province (with Wuhan as its capital), China for February.
- Relationships between policy, news or social sentiment changes, and outbreak severity are key to the longer-term control and constraining of the virus, and for human societies to manage to live with the virus, as many studies show that the virus will not be gone until a vaccine is developed to control the spreading [50]. These relationships are also critical for predicting the next wave of the outbreak and its epicenters as well as estimating risks for reopening the economy.
- Balancing the open data and privacy protection policies: Obtaining data is still quite challenging both nationally and internationally even NSF and the White House has issued open data policy with requirements for data management plan from funded projects [51]. More so for the pandemic research that engages a variety of geospatial location data from twitter location, location-based service and even medical records. These datasets involve strong privacy elements and should be protected while being used for research to fight against COVID-19 [52]. A platform, such as Geospatial Virtual Data Enclave (GVDE) [53], that can enable the sharing but at the same time protect the privacy is critical to safely archive, access, share, analyze, and use confidential geospatial data in COVID-19 research and other public health programs [53,54].

## 5. Discussion: Reflections regarding the impact on human society

The outbreak of COVID-19 reveals that humankind is not ready for a global pandemic and didn't learn enough from the outbreaks of SARS and MERS in past decades [13]. This pandemic drives us to prepare from many aspects, from psychological maturity for universal precautions, government bodies, employment and education, as well as nutrition, to improving our facilities and health-care infrastructure to become ready for the next emergency crises [55]. Even more challenging after the COVID -19 outbreak will be the reshaping of various aspects of human society.

### *5.1. The responding policies and administrative measures*

The COVID-19 triggered responses from many different domains and called for collaboration among all sectors to participate in research, engineering, production, delivery, and care, for rapid response to save lives and prepare communities. This outbreak also calls for a systematic study of the spatiotemporal topics in a controlled environment at a grand scale, such as pollution control, urban functions, global supply chains, and regional service location-allocation [55]. Such a cross-sector and all disciplinary mobilization calls for the whole human kind to fight against one enemy-the covid-19 disease, and is costing the world economy more than most regional wars or other natural disasters of the past century [56]. A collaboration based on global debates among different regions is needed and will be needed even more so after the fight against COVID-19. A cohesive and flexible framework with both spatial and temporal perspective in integrative fashion [57] will be a principal key to bridge interdisciplinary collaboration, cross-sector integration, and cross-region dialogues [58]. Spatiotemporal readiness for the next emergency is critical as evidenced by the promising signs and exposed weakness of the current status, such as many grassroot efforts, or the lack of standards and coordination.

### *5.2. Life on Earth*

Few events could have an impact on lives from around the globe at this scale. The pandemic is changing human lives and affecting many other species and ecosystems. Time will be needed to reach another equilibrium or relatively stable states in politics, culture, economics, and ecology. The lockdowns, closure of campuses, and stay-at-home orders while fighting against the pandemic also pose great and long-term challenges to education and impact the mental health of students and academic faculty and staff [59]. The consumption and working habits of many citizens have changed due to the stay-at-home policies. More and more people have become used to and are willing to purchase on-line and work from home. These phenomena will periodically or even permanently influence the economic structure and usage of public resources.

### *5.3. The resilience of a natural and artificial world*

The origin of the virus and whether it was leaked from a lab is a heated debate among many countries [60] and it brought a broader concern about how the artificial world should be in alliance with the natural world [61], so that scientists behind advanced research are equipped with humanitarian thinking and priority in mind when conducting any type of research [62]. The evolution of human knowledge and the advancement of technologies will put a lot of tools in the hands of decision makers, such as driverless cars, robotic personal service, and automatic battle machines. A debate is more pertinent on how to use such power to protect humankind. There are many ethical questions that remain unanswered as human beings seek answers to scientific questions, exploring and altering the micro and macro environments, and expanding our footprint not only on Earth but also in outer space.

### *5.4. Politics, humanity, and rapid response to global emergency*

It is a delicate balance in governance between saving lives and maintaining livelihood under the threat of this pandemic. The various response measures imposed by different countries have also generated heated debate on the effectiveness of governance models and organizational structures. Humanitarian activities in response to emergencies sometimes clash with the need for protecting sensitive information, privacy protection, and information classification. The existing political framework and communication channels may have to be reshaped if this global pandemic will be prolonged. Some argue that protecting human lives is more important than protecting human rights. Others urge for relaxing privacy protection in order to enable research, especially when there is an urgent need for fighting a global crisis and where individuals' locations are part of the critical data. Spatiotemporal and geospatial technology may provide a balanced solution for enabling effective use of location-based data while protecting citizen's privacy. For example, while using mobile phone data to track human movement sparks privacy concerns [63], developing novel spatiotemporal computing and aggregation algorithms could enable us to efficiently extract needed population flows from less sensitive but noisier social-media data.

### *5.5. An open world and global village*

The global village notion captured the rapid spreading and movement of the epicenter from Wuhan to Iran, Italy, Europe, New York, and now with significant outbreaks in Africa. The disease quickly spread across Africa and the time is limited for global leaders to control the transmission in Africa and back to other continents. The global framework of human societies is being tested and will very possibly be evaluated after the COVID-19 pandemic [8]. Haass [64] argued that the global pandemic may push world collaboration and contest back to the level of World War I or II, with greater disparities, though the advancement of human technologies and collaboration agreements have gone far beyond that of a century ago. While more discussions were towards a more transparent governance process, a more decisive mechanism for global emergency response may be formed, the deliberation and the process for moving ahead along those directions will probably be a compromise among different countries, cultures, and organizations. For sure it will take a long term to settle on a more effective mechanism for combating global disasters causing health crises such as COVID-19, natural disasters such as super typhoons or hurricanes [20], earthquakes [65], and tsunamis [66], as well as outer-space threats to our home planet [67]. Whatever the eventual outcome of this pandemic, it is clear that the world will be changed, and many relationships will be reshaped.

**Author Contributions:** C.Y. and W. G. came up with the original concept and coordinated the research initiative; S.B. contributed to the original research idea; C.Y. advised D.S., Q.L., Y.L., H.L. on the system designs and data collection, analytics; W.G. and S.B. advised T. H. on the data collection and analytics; D.S. coordinated the tasks; D.S., Y.T., Y.L. collaborated on collecting case data; D.S., C.Z., Y.T. developed the scripts for case data collection and cleaning, Q.L., and W.L. develop scripts for environmental data collection; C.Z., F.B., Y.T., K.C. collected the virus cases data and conducted quality control; H.L., Z.W., and H. L. provided computing support; C.Y., M.Y. and D.S. discussed and drew the spatiotemporal matrix; Z.Z., Y.L., D.S., conducted case spreading visual analytics; D.S., Y.L., Y.Z. conducted the policy stringency index visual analytics; Q.L., D.S., Y. L., and Y.Z. conducted the policy impact analytics; Q.L. conducted the remote sensing and night light analytics; Q.L., Z.Z, L.Z., M.L., and R.H. processed and visualized the air quality data; T. H., Z. L., Y.L., H.Z., and A. K. processed, analyzed and visualized the population movement; T.H., and D.S. analyzed the social implications; S.R. and D.S. visualized and analyzed the economic impact in the stock market; H.L., Y. L., Q. L. analyzed and conducted the simulations; C. Y. and all wrote the paper; W.G., D. W., D. P., M.Y., M. L., S.B., M.L., S.D., L.L., C.L., J.J., C.L., and D. R. discussed the idea and content; C.Y., W.G., D.W., D. R., Z.L., M.Y., L.Z., and D.P. revised the paper. All authors have read and agreed to the published version of the manuscript.

**Funding:** This research is supported by NSF CNS, ACI and BCS (1841520, 1835507 and 2025783), and the NSF Spatiotemporal Innovation Center members.

**Acknowledgments:** Jianjun Xu and Kyle Hawkins from Amazon helped apply for and utilize Amazon cloud resources, and we are thankful to our taskforce members. Dr. Michael Goodchild advised and helped review an earlier version of the paper.

**Conflicts of Interest:** The authors declare no conflict of interest.